\documentstyle[12pt]{article}

\begin{document}
\hfill\vbox{
\hbox{OHSTPY-HEP-T-95-020}
\hbox{hep-ph/9510408}
\hbox{October 1995} }\par
\thispagestyle{empty}

\begin{center}
{\Large \bf Free Energy of QCD at High Temperature}

\vspace{0.15in}

Eric Braaten and Agustin Nieto\\

{\it Department of Physics, Ohio State University,
Columbus, OH 43210, USA}

\end{center}

\begin{abstract}
Effective-field-theory methods are used to separate
the free energy for a nonabelian gauge theory at high temperature $T$
into the contributions from the momentum scales $T$, $gT$, and $g^2T$,
where $g$ is the coupling constant at the scale $2 \pi T$.
The effects of the scale $T$ enter through the coefficients in the
effective lagrangian for the 3-dimensional effective theory obtained
by dimensional reduction.  These coefficients can be calculated as
power series in  $g^2$.
The contribution to the free energy from the scale $gT$ can be calculated
using perturbative methods in the effective theory.  It can be expressed
as an expansion in $g$ starting at order $g^3$.
The contribution from the scale $g^2T$ must be calculated using
nonperturbative methods, but nevertheless
it can be expanded in powers of $g$ beginning
at order $g^6$.  We calculate the free energy explicitly to order $g^5$.
We also outline the calculations necessary to obtain the free energy
to order $g^6$.
\end{abstract}

\newpage
\section{Introduction}

One of the most dramatic predictions of quantum chromodynamics (QCD)
is that when hadronic matter is raised to a sufficiently high temperature
or density, it will undergo a phase transition to a quark-gluon plasma.
One of the major thrusts of nuclear physics in the next decade will
be the  effort to study the quark-gluon plasma through
relativistic heavy-ion collisions.  For this effort to be successful,
it will be important to understand the
properties of the plasma as accurately as possible.  The two major
theoretical tools that have been used to study the quark-gluon plasma
are lattice gauge theory and perturbative QCD.  Lattice gauge theory
has the advantage that it is a nonperturbative method and applies equally
well to the quark-gluon phase and to the hadron phase.
It is an effective method
for calculating the static equilibrium properties of hadronic matter with 0
baryon density.  Unfortunately, the Monte Carlo methods used in lattice
gauge theory can not be easily applied to problems involving dynamical
properties or to hadronic matter that is away from thermal equilibrium or
has nonzero baryon density.   These are severe restrictions, because a
quark-gluon plasma that is produced in heavy-ion collisions will not be at
thermal equilibrium and it may have nonzero baryon density.  Furthermore,
many of the most promising signatures for a quark-gluon plasma involve
dynamical properties.

Perturbative QCD can help fill this gap, at least for the quark-gluon phase
of hadronic matter.  This method can certainly be applied to the static
equilibrium properties of a quark-gluon plasma at 0 baryon density,
but there are no apparent obstacles to also applying it to dynamical problems,
or to non-equilibrium situations, or to a plasma with nonzero baryon density.
Thus it is a powerful tool for studying various aspects of
the quark-gluon plasma that might be probed through heavy-ion collisions.
However there are potential difficulties in applying perturbative QCD
to the quark-gluon plasma.
The method is based on treating the coupling constant $g$ as a small
parameter, but $g(\mu)$ is a parameter that varies rather dramatically with the
momentum scale $\mu$.  In order to apply perturbative QCD, it is
necessary  that $g$ be small at the
scale of the typical momentum of a particle in the plasma, which
is of order $T$ or perhaps $2 \pi T$.  While this is necesssary,
it may not be sufficient.  At sufficiently high
order in perturbation theory, any observable becomes sensitive to
low momentum gluons that interact with a large coupling strength $g$.
In order to rigorously apply perturbative QCD, it is essential to be
able to unravel the various momentum scales that play an important
role in a problem.  If low-momentum contributions are important,
they must be treated using nonperturbative methods.

For a quark-gluon plasma at high temperature, there is a hierarchy of 3
momentum scales that play an important role in static properties.
First, there is the scale $T$ of the typical momentum of a
particle in the plasma.  Next, there is the scale $gT$
associated with the screening of color-electric forces by the plasma.
Finally, there is the scale $g^2T$ associated with color-magnetic screening.
Only recently has a method been developed that can systematically
unravel the contributions from these various momentum scales.
The method is based on the construction of effective field theories that
reproduce static observables at successively longer distance scales.
This effective-field-theory approach is based on
an old idea called  ``dimensional reduction''
\cite{g-p-y,appelquist-pisarski}.  According to
this idea, the static properties of a (3+1)-dimensional field theory
at high temperature can be expressed in terms of an effective field
theory in 3 space dimensions.  Dimensional reduction has long been used
to provide insight into the qualitative behavior of field theories at
high temperature \cite{g-p-y,appelquist-pisarski,nadkarni-1,landsman}.
The effective-field-theory approach makes this idea into
a practical tool for quantitative calculations.  In Ref. \cite{eft},
we developed the effective-field-theory approach to dimensional
reduction and applied it to a scalar field with a $\phi^4$ interaction.
We demonstrated the power of this method by using it to carry out several
perturbative calculations beyond the frontiers set by previous work.
A similar approach was developed independently by Farakos, Kajantie,
Rummukainen, and Shaposhnikov \cite{f-k-r-s},
who applied it to the important problem of the
electroweak phase transition.  This method has also been applied to QCD
\cite{solution}, and used to resolve a longstanding  problem involving
the breakdown of the perturbation expansion for the free energy \cite{linde}.
These ideas have also been used to determine the asymptotic behavior of the
correlator of Polyakov loop operators \cite{polyakov} and to provide a rigorous
nonperturbative definition of the Debye screening mass in nonabelian
gauge theories \cite{arnold-yaffe}.

Once we have understood how to resolve the contributions of the various
momentum scales in thermal QCD, asymptotic freedom guarantees us that
perturbation theory will be under control in the high temperature limit.
At sufficiently high temperature, the running coupling constant will be
small enough that calculations to leading order in $g$ will be accurate.
However, in most practical applications, such as those encountered
in heavy ion collisions, the temperature is not asymptotically large,
and we must worry about higher order corrections.
The accuracy of the perturbation expansion can only be assessed
by carrying out explicit perturbative calculations beyond leading order.
One of the obstacles to progress in high temperature field theory
has been that the technology for perturbative calculations was not
well developed.  Only very recently have there been any calculations
to a high enough order that the running of the coupling
constant comes into play.  The simplest physical observable that
can be calculated in perturbation theory is the free energy,
which determines all the static thermodynamic properties of the system.
The running of the coupling constant first enters at order $g^4$.
The free energy for gauge theories at 0 temperature but large
chemical potential was calculated to order $g^4$ long ago \cite{mclerran}.
The first such calculation at high temperature was the free energy
of a scalar field theory with a $\phi^4$ interaction,
which was calculated to order $g^4$ by Frenkel, Saa, and Taylor in 1992
\cite{f-s-t}.  (A technical error was later corrected by Arnold and Zhai
\cite{arnold-zhai}.)  The analogous calculations for gauge theories
were carried out in 1994.  The free energy for QED was calculated
to order $e^4$ by Coriano and Parwani \cite{coriano-parwani} and the free
energy for a nonabelian gauge theory was calculated to order $g^4$ by
Arnold and Zhai \cite{arnold-zhai}.  The calculation of Arnold and Zhai
was completely analytic, and thus represent a particularly
significant leap in calculational technology.
The calculational frontier has since been extended to fifth order
in the coupling constant by Parwani and Singh
\cite{parwani-singh} and by Braaten and Nieto \cite{eft} for the
$\phi^4$ field theory,  by Parwani \cite{parwani} and by Andersen
\cite{andersen} for QED, and by
Kastening and Zhai \cite{kastening-zhai} for nonabelian gauge theories.
In this paper, we present an independent calculation of the free energy
for a nonabelian gauge theory to order $g^5$ \cite{fQCD},
verifing the result of Kastening and Zhai.
In our calculation, we use effective-field-theory
methods to simplify the calculation and to resolve the
contributions to the free energy from the momentum scales $T$ and $gT$.
We also outline the calculations that are required to obtain the free
energy to order $g^6$.

In Section 2, we describe how effective field theories can be used to
resolve the contributions to the free energy from the momentum scales
$T$, $gT$, and $g^2T$.  In Section 3, we calculate the coefficients in the
lagrangian for the effective field theory obtained by dimensional reduction.
In Section 4, we use the effective field theory to calculate the free energy
for QCD to order $g^5$.  In Section 5, we outline the
calculations that would be necessary to improve the accuracy to order $g^6$.
In Section 6, we discuss the implications of our calculation for
convergence of the perturbation expansion for the free energy.
We present some conclusions in  Section 7.
In two appendices, we tabulate the analytic expressions for all
the sums and integrals that arise in our calculation.

\section{Separation of Scales in the Free Energy}

The free energy for QCD at high temperature $T$ includes contributions
>from the momentum scales $T$, $gT$, and $g^2T$.  In this section,
we explain how the contributions from these three momentum scales can
be unraveled by using effective field theory methods.

The static equilibrium properties for hot QCD are given by
the free energy density $F$, which is proportional to the logarithm
of the partition function:
\begin{equation}
F \;=\; - {T \over V} \log {\cal Z}_{\rm QCD} \;,
\end{equation}
where $V$ is the volume of space.
In the imaginary-time formalism for thermal QCD, the partition function
is given by a functional integral over quark and gluon fields
on a 4-dimensional Euclidean space.  The Euclidean time $\tau$
is periodic with period $\beta = 1/T$.   The partition function is
\begin{equation}
{\cal Z}_{\rm QCD} \;=\;
\int {\cal D} A_\mu({\bf x},\tau) {\cal D} q({\bf x},\tau)
	{\cal D} \bar q({\bf x},\tau)
\exp \left( - \int_0^\beta d \tau \int d^3x \; {\cal L}_{\rm QCD} \right) \; .
\label{ZQCD}
\end{equation}
The gluon fields are periodic functions of $\tau$ while the quark and
antiquark fields are antiperiodic.  The lagrangian is
\begin{equation}
{\cal L}_{\rm QCD} \;=\; {1 \over 4} G^a_{\mu \nu} G^a_{\mu \nu}
\;+\; \bar q \gamma_\mu D_\mu q \;,
\label{LQCD}
\end{equation}
where $G^a_{\mu \nu} = \partial_\mu A^a_\nu - \partial_\nu A^a_\mu
	+ g f^{abc} A^b_\mu A^c_\nu$
is the field strength and $g$ is the gauge coupling constant.
All the quark fields have been assembled into the multi-component
spinor $q$,
and the gauge-covariant derivative acting on this spinor is
$D_\mu = \partial_\mu + i g A_\mu^a T^a$.
The relevant quark flavors are all assumed to be massless.

In order to make our calculations as general as possible, we will express
them in terms of the group-theory factors $C_A$, $C_F$, and $T_F$
defined by
\begin{eqnarray}
f^{abc} f^{abd} &=& C_A \delta^{cd} \;,
\\
\left( T^a T^a \right)_{ij} &=& C_F \delta_{ij} \;,
\\
{\rm tr} \left( T^a T^b \right) &=& T_F \delta^{ab} \;.
\end{eqnarray}
For an $SU(N_c)$ gauge theory with $n_f$ quarks in the fundamental
representation, these factors are
$C_A = N_c$, $C_F = (N_c^2 - 1)/(2 N_c)$, and $T_F = n_f/2$.
The dimensions of the adjoint representation and the fermion
representation are $d_A = N_c^2 - 1$ and $d_F = N_c n_f$, respectively.

The free energy for QCD can also be calculated using an effective field theory
in 3 space dimensions called electrostatic QCD (EQCD).
This effective theory is constructed
so that it reproduces static gauge-invariant
correlators of QCD at distances of order $1/(gT)$ or larger.
It contains an electrostatic
gauge field $A_0^a({\bf x})$ and a magnetostatic gauge field $A_i^a({\bf x})$
that can be identified, up to field redefinitions,
with the zero-frequency modes of the gluon field $A^a_\mu({\bf x}, \tau)$
for thermal QCD in a static gauge \cite{nadkarni-1}.
The free energy for thermal QCD can be written
\begin{equation}
F \;=\; T \; \left( f_E(\Lambda_E)
	\;-\; {\log {\cal Z}_{\rm EQCD} \over V} \right)\,,
\label{FEQCD}
\end{equation}
where ${\cal Z}_{\rm EQCD}$ is the partition function for EQCD:
\begin{equation}
{\cal Z}_{\rm EQCD} \;=\;
\int^{(\Lambda_E)} {\cal D} A_0({\bf x}) {\cal D} A_i({\bf x})
\exp \left( - \int d^3x \; {\cal L}_{\rm EQCD} \right).
\label{ZEQCD}
\end{equation}
The functional integral requires an ultraviolet cutoff $\Lambda_E$.
The lagrangian for EQCD is
\begin{equation}
{\cal L}_{\rm EQCD} \;=\;
{1 \over 4} G^a_{ij} G^a_{ij} \;+\; {1 \over 2} (D_i A_0)^a (D_i A_0)^a
\;+\; {1 \over 2} m_E^2 A_0^a A_0^a
\;+\; {1 \over 8} \lambda_E (A_0^a A_0^a)^2
	\;+\; \delta {\cal L}_{\rm EQCD},
\label{LEQCD}
\end{equation}
where $G^a_{ij} = \partial_i A^a_j - \partial_j A^a_i
	+ g_E f^{abc} A^b_i A^c_j$
is the magnetostatic field strength with coupling constant
$g_E$.
If the fields $A_0$ and $A_i$ are assigned the scaling
dimension $1/2$, then the operators shown explicitly in (\ref{LEQCD})
have dimensions 3, 3, 1, and 2, respectively.
The term $\delta {\cal L}_{\rm EQCD}$
in (\ref{LEQCD}) includes all other local gauge-invariant operators
of dimension 3 and higher that can be constructed out of $A_0$ and $A_i$.
Static gauge-invariant correlation functions in full QCD
can be reproduced in EQCD
by tuning the gauge coupling constant $g_E$, the mass parameter $m_E^2$,
the coupling constant $\lambda_E$,
and the parameters in $\delta {\cal L}_{\rm EQCD}$ as functions of
$g$, $T$ and the ultraviolet cutoff $\Lambda_E$ of EQCD.
The $\Lambda_E$-dependence of the parameters is cancelled by the
$\Lambda_E$-dependence of the loop integrals in the effective theory.

In order to calculate the free energy using EQCD, we must also tune the
coefficient $f_E$ of the unit operator, which was omitted from
the effective lagrangian (\ref{LEQCD}) but appears
as the first term in the expression (\ref{FEQCD}) for the free energy.
It depends on the ultraviolet cutoff
$\Lambda_E$ of EQCD in such a way as to cancel the
cutoff dependence of the partition function for EQCD.
The coefficient $f_E$ gives the contribution to the free energy from
the momentum scale $T$. The logarithm of the partition function
for EQCD includes the remaining contributions from the smaller
momentum scales $gT$ and $g^2T$.

In order to further separate the contributions from the scales
$gT$ and $g^2T$, it is convenient to construct a second effective
field theory called magnetostatic QCD (MQCD) which contains
only the magnetostatic gauge field $A^a_i({\bf x})$.
The free energy  for thermal QCD can be written
\begin{equation}
F \;=\; T \; \left( f_E(\Lambda_E) \;+\; f_M(\Lambda_E,\Lambda_M)
	\;-\; {\log {\cal Z}_{\rm MQCD} \over V} \right)\,,
\label{FMQCD}
\end{equation}
where ${\cal Z}_{\rm MQCD}$ is the partition function for MQCD:
\begin{equation}
{\cal Z}_{\rm MQCD} \;=\;
\int^{(\Lambda_M)} {\cal D} A_i^a({\bf x})
\exp \left( - \int d^3x \; {\cal L}_{\rm MQCD} \right).
\label{ZMQCD}
\end{equation}
The functional integral requires an ultraviolet cutoff $\Lambda_M$.
The lagrangian for MQCD is
\begin{equation}
{\cal L}_{\rm MQCD} \;=\;
{1 \over 4} G^a_{ij} G^a_{ij}
\;+\; \delta {\cal L}_{\rm MQCD} ,
\label{LMQCD}
\end{equation}
where $G^a_{ij}$ is the magnetostatic field strength with coupling constant
$g_M$. This coupling constant differs from $g_E$ by perturbative corrections.
The term $\delta {\cal L}_{\rm MQCD}$ includes all possible local
gauge-invariant operators of dimension 5 and higher
that can be constructed out of $A_i^a$.
Gauge-invariant correlation functions in EQCD
can be reproduced in MQCD
by tuning the gauge coupling constant $g_M$
and the parameters in $\delta {\cal L}_{\rm MQCD}$ as functions of
the parameters of EQCD ($g_E$, $m_E^2$, $\lambda_E$, $\ldots$)
and the ultraviolet cutoff $\Lambda_M$ of MQCD.
The $\Lambda_M$-dependence of the parameters in the MQCD lagrangian
is cancelled by the $\Lambda_M$-dependence of the loop integrals in MQCD.

In order to calculate the free energy using MQCD, one must also tune
the coefficient $f_M$ of the unit operator,
which was omitted from the effective lagrangian (\ref{LMQCD})
but appears as the second term in the expression (\ref{FMQCD})
for the free energy.
Its dependence on  the ultraviolet cutoff $\Lambda_M$ of MQCD
is cancelled by the
cutoff dependence of the partition function for MQCD.
The coefficient $f_M$ gives the contribution to the free energy from the
momentum scale $gT$.  The contribution from the smaller momentum scale
$g^2T$ is contained in the logarithm of the partition function for
MQCD.

By constructing the effective field theories EQCD and MQCD,
we have separated the contributions from the momentum scales
$T$, $gT$, and $g^2T$ in the free energy.  The general structure of
the free energy is
\begin{equation}
F \;=\;  T \; \left[ f_E(T,g;\Lambda_E)
\;+\; f_M(m^2_E,g_E,\lambda_E,\ldots;\Lambda_E,\Lambda_M)
	\;+\; f_G(g_M,\ldots;\Lambda_M) \right] T \;,
\label{F123}
\end{equation}
where $f_G = - \log {\cal Z}_{\rm MQCD}/V$.
The arbitrary factorization scales $\Lambda_E$ and $\Lambda_M$
separate the momentum scales $T$ from $gT$ and $gT$ from $g^2T$,
respectively.
The term $f_E$ and the parameters of EQCD
(i.e., $m^2_E$, $g_E$, $\lambda_E$, $\ldots$)
involve only the scale $T$.  They can therefore be calculated
using ordinary perturbation theory
as power series in $g^2(2 \pi T)$, where $g(2 \pi T)$ is the running coupling
constant at the scale of the lowest Matsubara frequency $2 \pi T$.
The term $f_M$ and the parameters of MQCD ($g_M$, $\ldots$)
involve only the scale $g T$.  They can be calculated
in EQCD as perturbation expansions in $g_E^2/m_E$, $\lambda_E/m_E$,
and other dimensionless parameters obtained by multiplying
EQCD coupling constants by appropriate powers of $m_E$.
The leading contribution to $f_M$ is proportional to $m_E^3$.
The term $f_G$ in (\ref{F123}) can only be calculated using
nonperturbative methods, such as lattice gauge theory simulations of MQCD.
Surprisingly, however, $f_G$ can be expanded as a weak coupling
expansion in powers of $g$ by treating the higher dimension operators
in the MQCD lagrangian as perturbations \cite{solution}.
The leading term is proportional to $g_M^6$.

In summary, the free energy for QCD has the general structure given in
(\ref{F123}).  The term $f_E$ is the contribution from the scale $T$.
It has the form $T^3$ multiplied by
a power series in $g^2(2 \pi T)$ whose coefficients can be calculated
using ordinary perturbation theory in thermal QCD.
The term $f_M$ is the contribution from the scale $gT$.  It has the
form $m_E^3$ multiplied by
a power series in $g(2 \pi T)$ whose coefficients
can be calculated using perturbation theory in EQCD.
The term $f_G$ is the contribution from the scale $g^2T$.
It has the form $g_M^6$ multiplied by a power series
in $g(2 \pi T)$ whose coefficients can be calculated using
lattice simulations of MQCD.

\section{Parameters in the EQCD Lagrangian}

In order to calculate the free energy using the EQCD lagrangian,
the parameters in the lagrangian (\ref{LEQCD}) must be tuned
as functions of $g$, $T$, and $\Lambda_E$ so that EQCD
reproduces the static gauge-invariant
correlation functions of full QCD at distances
$R \gg 1/T$.  The EQCD parameters can be determined by computing various
static quantities in full QCD, computing the corresponding
quantities in EQCD, and demanding that they match.
It is convenient to carry out these matching calculations
using a strict perturbation expansion in $g^2$.
This expansion is afflicted with infrared divergences.
The divergences arise from long-range forces mediated by static gluons,
which remain massless in the strict perturbation expansion.
Physically, these divergences are screened by plasma effects either
at the scale $gT$ in the case of electrostatic gluons
or at the scale $g^2 T$ in the case of magnetostatic gluons.
The screening of electrostatic gluons can be taken into account by
summing up infinite sets of higher-order diagrams in the perturbation
expansion, but the screening of magnetostatic gluons can only be
taken into account using nonperturbative methods.
Fortunately, it is not necessary to treat the effects of screening
in a physically correct way in order to determine the parameters
in the EQCD lagrangian. The parameters take into account the effects
of large momenta of order $T$, and they are therefore insensitive to the
infrared effects associated with screening.  We can therefore
simply ignore screening and remove the infrared divergences in the
strict perturbation expansion by imposing any convenient infrared cutoff.
As long as we use the same infrared cutoff in EQCD and in
full QCD, we can determine the EQCD parameters by matching strict
perturbation expansions in the two theories.
Note that we are using the strict perturbation expansion simply as
a device for determining the parameters in the EQCD lagrangian.

\subsection{Gauge coupling constant}

For the calculation of the free energy to order $g^5$,
we require the EQCD gauge coupling constant $g_E$ only to
leading order in $g^2$.  At this order, we can simply read $g_E$ off
>from the lagrangian of the full theory.
We substitute $A_0({\bf x},\tau) \longrightarrow \sqrt{T} A_0({\bf x})$ in
the QCD lagrangian (\ref{LQCD})
and compare $\int_0^\beta d \tau {\cal L_{\rm QCD}}$ with
${\cal L}_{\rm EQCD}$ in (\ref{LEQCD}).  We find that, to leading order
in $g^2$,
\begin{equation}
g_E^2 \;=\; g^2 T .
\label{gE}
\end{equation}
There is no dependence on the factorization scale $\Lambda_E$ at this order.
The coupling constant $g_E$ could be calculated to higher order
in $g^2$ by matching scattering amplitudes
in full QCD with the corresponding ones in EQCD.

\subsection{Mass parameter}

In this subsection, we calculate the coefficient $m_E^2$
of the $A^a_0 A^a_0$ term in the EQCD lagrangian
to next-to-leading order in $g^2$.  The physical interpretation
of $m_E$ is that it is  the contribution to the electric screening
mass $m_{\rm el}$ from large momenta of order $T$.
The parameter $m_E^2$ can be
determined by matching the strict perturbation expansions for the
electric screening mass in full QCD and in EQCD.
Beyond leading order in $g$, the electric screening
mass becomes sensitive to magnetostatic screening
and requires a nonperturbative definition \cite{arnold-yaffe}.
However, in the presence of an infrared cutoff, $m_{\rm el}$
can be defined in full QCD by the condition that the propagator
for the field $A_0^a(\tau,{\bf x})$ at spacelike momentum
$K = (k_0=0,{\bf k})$ has a pole at ${\bf k}^2 = - m_{\rm el}^2$.
It is the solution to the equation
\begin{equation}
k^2 \;+\; \Pi(k^2) \;=\; 0
\qquad \mbox{at $k^2=-m_{\rm el}^2$},
\label{msdef}
\end{equation}
where $\Pi(k^2)$ is the $\mu=\nu=0$ component of the gluon self-energy
tensor evaluated at $k_0=0$:
$\Pi^{ab}_{00}(k_0=0,{\bf k}) = \Pi(k^2) \delta^{ab}$.
In EQCD with an infrared cutoff, the electric screening mass
$m_{\rm el}$ gives the location of the pole in the propagator
for the field $A_0^a({\bf x})$.  Denoting the self-energy function
by $\Pi_E(k^2) \delta^{ab}$, $m_{\rm el}$ is the solution to
\begin{equation}
k^2 \;+\; m_E^2 \;+\; \Pi_E(k^2) \;=\;0
\qquad \mbox{at $k^2=-m_{\rm el}^2$}.
\label{msdefeff}
\end{equation}
By matching the expressions for $m_{\rm el}$
obtained by solving (\ref{msdef}) and (\ref{msdefeff}),
we can determine the parameter $m_E^2$.

We calculate the electric mass $m_{\rm el}$ in the full theory using a
strict perturbation expansion in $g^2$ and using dimensional regularization
with $3-2 \epsilon$ spacial dimensions to cut off both infrared and
ultraviolet divergences.
The self-energy function $\Pi(k^2)$ can be expanded in a
loop expansion
\begin{equation}
\Pi(k^2)
\;=\; \Pi^{(1)}(k^2) \;+\; \Pi^{(2)}(k^2) \;+\; \ldots \;,
\end{equation}
with $\Pi^{(1)}(k^2)$ and $\Pi^{(2)}(k^2)$
being given by the diagrams in Fig.~1 and 2, respectively.
We can simplify the equation (\ref{msdef}) by expanding $\Pi(k^2)$
as a Taylor expansion around $k^2 = 0$.  This is justified by the
fact that the leading order solution to (\ref{msdef}) gives a value
of $k^2$ that is of order $g^2 T^2$.  The deviation of $k^2$
>from 0 should therefore be treated as a perturbation in order
to get the strict perturbation expansion for $m_{\rm el}^2$
in powers of $g^2$.  The resulting expression for the electric
screening mass to next-to-leading order in $g^2$ is
\begin{equation}
m_{\rm el}^2 \;\approx\;
\Pi^{(1)}(0) \;+\; \Pi^{(2)}(0)
\;-\; \Pi^{(1)}(0) {d \Pi^{(1)} \over d k^2}(0) \;.
\label{melsq}
\end{equation}
Here and below, we use the symbol ``$\approx$'' to denote an
equality that holds only in the strict perturbation expansion.
The one-loop diagrams that contribute to $\Pi^{(1)}(k^2)$
are shown in Fig.~1.
Evaluating this function and its first derivative
at $k^2 = 0$ in Feynman gauge, we obtain
\begin{eqnarray}
\Pi^{(1)}(0) &\approx& Z_g^2 g^2  \;
\Bigg\{ 2 (1 - \epsilon) C_A ( {\cal I}_1 - 2 {\cal J}_1 )
\;-\; 4 T_F ( \widetilde{\cal I}_1 - 2 \widetilde{\cal J}_1 ) \Bigg\} \;,
\label{Pi10}
\\
{d \Pi^{(1)} \over d k^2}(0) &\approx&
g^2 \Bigg\{
- 2 C_A \left[ {\cal I}_2
	+ {2(1-\epsilon)(1+2\epsilon) \over 3-2\epsilon} {\cal J}_2
	- {8(1-\epsilon) \over 3-2\epsilon} {\cal K}_2 \right]
\nonumber \\
&& \;+\; 2 T_F \left[ \widetilde{\cal I}_2
	+ {4(1+2\epsilon) \over 3-2\epsilon} \widetilde{\cal J}_2
	- {16 \over 3-2\epsilon} \widetilde{\cal K}_2 \right]
\Bigg\} .
\label{dPi10}
\end{eqnarray}
The sum-integrals ${\cal I}_n$, ${\cal J}_n$, ${\cal K}_n$
$\widetilde{\cal I}_n$, $\widetilde{\cal J}_n$, and $\widetilde{\cal K}_n$
are defined in appendix A. The renormalization of the coupling constant
using the $\overline{\rm MS}$ scheme is accomplished by substituting
\begin{equation}
Z_g^2 = 1
\;-\; {11 C_A - 4 T_F \over 3} {g^2 \over (4 \pi)^2}
	{1 \over \epsilon}
\label{Zg}
\end{equation}
into the expression for $\Pi^{(1)}(0)$.
The two-loop diagrams that contribute to $\Pi(k^2)$
are shown in Fig.~2.  This function evaluated at $k^2 = 0$ is
\begin{eqnarray}
\Pi^{(2)}(0) &\approx&
g^4  \Bigg\{
4 (1-\epsilon) C_A^2 \left[ - 2 {\cal I}_2 {\cal J}_1
	+ \epsilon {\cal I}_1 {\cal I}_2
	+ 4 (1-\epsilon) {\cal I}_1 {\cal J}_2 \right]
\nonumber \\
&& \;+\;
8 C_A T_F \left[ 2 {\cal I}_2 \widetilde{\cal J}_1
	- \epsilon \widetilde{\cal I}_1 {\cal I}_2
	- 4 (1-\epsilon) \widetilde{\cal I}_1 {\cal J}_2 \right]
\nonumber \\
&& \;+\;
8 (1-\epsilon) C_F T_F \left( {\cal I}_1 - \widetilde{\cal I}_1 \right)
	\left( \widetilde{\cal I}_2 - 4 \widetilde{\cal J}_2 \right)
\Bigg\}.
\label{Pi20}
\end{eqnarray}
The sum-integrals in (\ref{Pi10}), (\ref{dPi10}), and (\ref{Pi20})
can be evaluated analytically using methods developed by
Arnold and Zhai \cite{arnold-zhai}, and they are given in Appendix A.
The three quantities appearing in (\ref{melsq}) reduce to
\begin{eqnarray}
\Pi^{(1)}(0) &\approx& Z_g^2 g^2 \; T^2
\Bigg\{ {1 \over 3} C_A \left[ 1 \;+\; \left( 2 {\zeta'(-1) \over \zeta(-1)}
	+ 2 \log {\Lambda \over 4 \pi T} \right) \epsilon \right]
\nonumber \\
&& \;+\; {1 \over 3}T_F \left[ 1 \;+\; \left( 1 - 2 \log 2
	+ 2 {\zeta'(-1) \over \zeta(-1)}
	+ 2 \log {\Lambda \over 4 \pi T} \right) \epsilon \right]
\Bigg\} \;,
\\
{d \Pi^{(1)} \over d k^2}(0) &\approx&
{g^2 \over (4 \pi)^2}
\Bigg\{ - {5 \over 3} C_A \left[ {1 \over \epsilon} - {1 \over 5}
	+ 2 \gamma + 2 \log {\Lambda \over 4 \pi T} \right]
\nonumber \\
&& \;+\; {4 \over 3} T_F \left[ {1 \over \epsilon} - 1
	+ 4 \log 2 + 2 \gamma + 2 \log {\Lambda \over 4 \pi T} \right]
\Bigg\} \;,
\\
\Pi^{(2)}(0) &\approx&
{g^4 \over (4 \pi)^2}  \; T^2
\Bigg\{ {2 \over 3} C_A^2 \left[ {1 \over \epsilon} + 1 + 2 \gamma
	+ 2 {\zeta'(-1) \over \zeta(-1)} + 4 \log {\Lambda \over 4 \pi T}
	\right]
\nonumber \\
&& \;+\; {2 \over 3} C_A T_F
\left[ {1 \over \epsilon} + 2 - 2 \log 2 + 2 \gamma
	+ 2 {\zeta'(-1) \over \zeta(-1)} + 4 \log {\Lambda \over 4 \pi T}
	\right]
\;-\; 2 C_F T_F \Bigg\} \;,
\end{eqnarray}
where $\gamma$ is Euler's constant, $\zeta(z)$ is the Riemann zeta function,
and $\Lambda$ is the scale of dimensional regularization.
Inserting these expressions into (\ref{melsq}),
we find that the strict perturbation expansion for $m_{\rm el}^2$
to order $g^4$ is
\begin{eqnarray}
m_{\rm el}^2 &\approx&
{1 \over 3} g^2(\Lambda) T^2
\Bigg\{
C_A + T_F
\nonumber \\
&& \;+\; \epsilon \left[
C_A \left( 2 {\zeta'(-1) \over \zeta(-1)}
	+ 2 \log {\Lambda \over 4 \pi T} \right)
\;+\; T_F \left( 1 - 2 \log 2 + 2 {\zeta'(-1) \over \zeta(-1)}
	+ 2 \log {\Lambda \over 4 \pi T} \right)
\right]
\nonumber \\
&& \;+\; \left[
C_A^2 \left( {5 \over 3} + {22 \over 3} \gamma
	+ {22 \over 3} \log {\Lambda \over 4 \pi T} \right)
\;+\; C_A T_F \left( 3 - {16 \over 3} \log 2 + {14 \over 3} \gamma
	+ {14 \over 3} \log{\Lambda \over 4 \pi T} \right)
\right.
\nonumber \\
&& \left.
\;+\; T_F^2 \left( {4 \over 3} - {16 \over 3} \log 2 - {8 \over 3} \gamma
	- {8 \over 3} \log{\Lambda \over 4 \pi T} \right)
\;-\; 6 C_F T_F \right] {g^2 \over (4 \pi)^2}
\Bigg\} \;.
\label{melpert}
\end{eqnarray}
Note that all the poles in $\epsilon$ have cancelled.
In the order $g^2$ term, we have kept terms of order $\epsilon$
for later use.
The expression (\ref{melpert}) depends on $\Lambda$ explicitly
through logarithms of $\Lambda/4 \pi T$
and implicitly through the coupling constant $g^2(\Lambda)$.
The scale of the coupling constant can be shifted from the dimensional
regularization scale $\Lambda$ to an arbitrary renormalization scale $\mu$
by using the solution to the renormalization group
equation for the running coupling constant:
\begin{equation}
g^2(\Lambda) \;=\; g^2(\mu)
\left[ 1 \;+\; {2(11 C_A - 4 T_F) \over 3}  {g^2 \over (4 \pi)^2}
		\log {\mu \over \Lambda} \right] .
\label{glam}
\end{equation}
After making this shift in the scale of the coupling constant,
the only remaining dependence on $\Lambda$ occurs in the terms of
order $\epsilon$.  In these terms, $\Lambda$ can be identified with the
factorization scale $\Lambda_E$ that separates the scales $T$ and $gT$.

The expression (\ref{melpert}) for $m_{\rm el}^2$ is an expansion
in powers of $g^2$.  It does not include a $g^3$ term,
in contrast to the expression for
$m_{\rm el}^2$ that correctly incorporates the effects of the
screening of electrostatic gluons \cite{rebhan}.
This $g^3$ term arises because the $g^4$ correction includes a linear
infrared divergence that is cut off at the scale $gT$.
Since we have used dimensional regularization as an infrared cutoff,
power infrared divergences such as this linear divergence
have been set equal to 0.

In order to match with the expression (\ref{melpert}), we have to
calculate the screening mass in EQCD using the strict
expansion in $g^2$.  Since $m_E^2$ is treated as a perturbation
parameter of order $g^2$,
the only scale in the self-energy function $\Pi_E(k^2)$
is $k^2$.  After Taylor expanding in powers of $k^2$, there
is no scale in the dimensionally regularized integrals, so they all vanish.
The solution to the equation (\ref{msdefeff})
for the screening mass is therefore trivial:
\begin{equation}
m_{\rm el}^2 \;\approx\; m_E^2 .
\label{msperteff}
\end{equation}

Comparing (\ref{melpert}) and (\ref{msperteff}), we find that,
in the limit $\epsilon \to 0$,
the parameter $m_E^2$ is given by
\begin{eqnarray}
m_E^2 \Bigg|_{\epsilon = 0}
&=&  {1 \over 3} \; g^2(\mu) \; T^2
\Bigg\{ C_A + T_F
\nonumber \\
&& \;+\; \left[
C_A^2 \left( {5 \over 3} + {22 \over 3} \gamma
	+ {22 \over 3} \log {\mu \over 4 \pi T} \right)
\;+\; C_A T_F \left( 3 - {16 \over 3} \log 2 + {14 \over 3} \gamma
	+ {14 \over 3} \log{\mu \over 4 \pi T} \right)
\right.
\nonumber \\
&& \left.
\;+\; T_F^2 \left( {4 \over 3}- {16 \over 3} \log 2 - {8 \over 3} \gamma
	- {8 \over 3} \log{\mu \over 4 \pi T} \right)
\;-\; 6 C_F T_F \right]
	 {g^2 \over (4 \pi)^2}
\Bigg\} \;.
\label{mE}
\end{eqnarray}
At this order in $g^2$, there is no dependence on the
factorization scale $\Lambda_E$.
The order-$\epsilon$ terms in $m_E^2$ will also be required later in
the calculation.  These terms are
\begin{eqnarray}
{\partial m_E^2 \over \partial \epsilon} \Bigg|_{\epsilon = 0}
&=& {1 \over 3} g^2 T^2 \Bigg\{
C_A \left( 2 {\zeta'(-1) \over \zeta(-1)}
	+ 2 \log {\Lambda_E \over 4 \pi T} \right)
\nonumber \\
&& \;+\; T_F \left( 1 - 2 \log 2 + 2 {\zeta'(-1) \over \zeta(-1)}
	+ 2 \log {\Lambda_E \over 4 \pi T} \right) \Bigg\} .
\label{dmEdeps}
\end{eqnarray}
This expression depends explicitly on the factorization scale $\Lambda_E$.

\subsection{Coefficient of the unit operator}

In this subsection, we calculate the coefficient of the unit operator
$f_E$ to next-to-next-to-leading order in $g^2$. The physical
interpretation of $f_E$ is that $f_E T$
is the contribution to the free energy from large momenta of order $T$.
The parameter $f_E$
is determined by calculating the free energy
as a strict perturbation in $g^2$ in both full QCD and EQCD,
and matching the two results.

In the full theory, the free energy has a diagrammatic expansion
that begins with the one-loop, two-loop and three-loop diagrams
shown in Fig.~3, 4, and 5.  Evaluating the diagrams in Feynman gauge,
we obtain
\begin{eqnarray}
F &\approx&
- (1 - \epsilon) d_A {\cal I}'_0
\;+\; 2 d_F  \widetilde{\cal I}'_0
\nonumber \\
&& \;+\; d_A Z_g^2 g^2
\left[ (1 - \epsilon)^2 C_A {\cal I}_1^2
	+ 2(1-\epsilon)T_F \widetilde{\cal I}_1(\widetilde{\cal I}_1
               - 2{\cal I}_1) \right]
\nonumber \\
&&
\;+\; d_A g^4 \Bigg\{
C_A^2 (1-\epsilon)^2 \left[
  2 (1+\epsilon) {\cal I}_1^2 {\cal I}_2
  - {1\over 2} {\cal M}_{0,0}
  - {\cal M}_{2,-2}
\right]
\nonumber \\
&&
\qquad
\;+\; C_A T_F (1-\epsilon) \left[
  -8 {\cal I}_1 {\cal I}_2 \widetilde{\cal I}_1
  -2 \epsilon \widetilde{\cal M}_{0,0}
  + (1+\epsilon) {\cal N}_{0,0}
  + 4 \widetilde{\cal M}_{-2,2}
\right]
\nonumber \\
&&
\qquad
\;+\; T_F^2 \left[
  8 (1+\epsilon) {\cal I}_2 \widetilde{\cal I}_1^2
   + 2 \epsilon {\cal N}_{0,0}
   - 4 {\cal N}_{2,-2}
\right]
\nonumber \\
&&
\qquad
\;+\; 2 C_F T_F (1-\epsilon) \left[
  2 (1-\epsilon) \left( {\cal I}_1^2 - 4 {\cal I}_1 \widetilde{\cal I}_1
   + \widetilde{\cal I}_1^2 \right) \widetilde{\cal I}_2
  + 2 \widetilde{\cal M}_{0,0}
  - (1+\epsilon) {\cal N}_{0,0}
  \right.
  \nonumber \\
 && \left.
 \qquad \qquad \qquad \qquad \qquad \qquad
 + 2 (1-\epsilon) \widetilde{\cal M}_{1,-1}
\right]
\Bigg\} \,.
\label{Ffullint}
\end{eqnarray}
The symbol ``$\approx$'' is a reminder that the strict perturbation
in $g^2$ does not give a physically correct treatment of
the screening effects of the plasma.
The sum-integrals in (\ref{Ffullint}) are given in
Appendix~\ref{ap:a}. To order
$g^4$, the renormalization of the coupling constant is accomplished
in the $\overline{\rm MS}$ scheme by substituting (\ref{Zg})
for $Z_g$ in the order $g^2$ term.
The final result is
\begin{eqnarray}
F &\approx& - {\pi^2 d_A\over 9} T^4
\Bigg\{
{1 \over 5} \;+\; {7 \over 20} {d_f \over d_A}
\;-\; \left( C_A + {5 \over 2} T_F \right) {g^2(\Lambda) \over (4 \pi)^2}
\nonumber \\
&& \;+\;
\left[ C_A^2
\left( {12 \over \epsilon}
	+ {194 \over 3} \log{\Lambda \over 4\pi T}
	+ {116 \over 5}
	+ 4 \gamma + {220 \over 3} {\zeta'(-1) \over \zeta(-1)}
	- {38 \over 3} {\zeta'(-3) \over \zeta(-3)} \right)
\right.
\nonumber \\
&& \left.
\;+\; C_A T_F
\left( {12 \over \epsilon}
	+ {169 \over 3} \log{\Lambda \over 4\pi T}
	+ {1121 \over 60} - {157 \over 5} \log 2
	+ 8 \gamma + {146 \over 3} {\zeta'(-1) \over \zeta(-1)}
	- {1\over 3} {\zeta'(-3) \over \zeta(-3)} \right)
\right.
\nonumber \\
&& \left.
\;+\; T_F^2
\left( {20 \over 3} \log{\Lambda \over 4\pi T}
	+ {1 \over 3} - {88 \over 5} \log 2
	+ 4 \gamma + {16 \over 3} {\zeta'(-1) \over \zeta(-1)}
	- {8 \over 3} {\zeta'(-3) \over \zeta(-3)} \right)
\right.
\nonumber \\
&& \left.
\;+\; C_F T_F
\left( {105 \over 4} - 24 \log 2 \right) \right]
	\left( {g^2 \over (4 \pi)^2} \right)^2
\Bigg\} .
\label{Fpert}
\end{eqnarray}

In EQCD, the free energy is given by the expression (\ref{FEQCD}).
We calculate $\log {\cal Z}_{\rm EQCD}$ using the strict perturbation expansion
in which $g_E^2$ and $m_E^2$ are treated as perturbation parameters and
both infrared and ultraviolet
divergences are regularized using dimensional regularization.
Since diagrams with massless propagators and with no external
legs vanish in dimensional regularization, the only contribution to
$\log {\cal Z}_{\rm EQCD}$ which does not vanish comes from the counterterm
$\delta f_E$ which cancels ultraviolet divergences proportional to the
unit operator.  The resulting expression for the free energy is simply
\begin{equation}
F \; \approx \; ( f_E + \delta f_E ) \; T \;.
\label{Feff}
\end{equation}
The counterterm can be determined by calculating the ultraviolet
divergences in $\log {\cal Z}_{\rm EQCD}$.
If we use dimensional regularization together with a
minimal subtraction renormalization scheme in the effective theory,
then $\delta f_E$ is a polynomial in $g_E^2$, $m_E^2$,
and the other parameters in the lagrangian for EQCD.
The only combination of parameters that has dimension 3
and is of order $g^4$ is $g_E^2 m_E^2$.  Thus the leading term in
$\delta f_E$ is proportional to $g_E^2 m_E^2$.  The coefficient
is determined by a 2-loop calculation that is a trivial  part
of the 3-loop calculation in Section 4.  The result for the counterterm is
\begin{equation}
\delta f_E \;=\;
- {d_A C_A \over 4 (4 \pi)^2} g_E^2 m_E^2 {1 \over \epsilon} \;.
\label{deltafE}
\end{equation}
When expressing this counterterm in terms of the parameters $g$ and $T$
of the full theory, we must take into account the fact that $m^2_E$
multiplies a pole in $\epsilon$.  Thus in addition to expression for $m_E^2$
given in (\ref{mE}), we must also include the terms of order
$\epsilon$ which are given by (\ref{dmEdeps}).  The counterterm
(\ref{deltafE}) is therefore
\begin{eqnarray}
\delta f_E \;=\;
- {\pi^2 d_A \over 9}
	\left( {g^2 \over (4 \pi)^2} \right)^2 T^3 \;
\left[
12 C_A^2 \left( {1 \over \epsilon} + 2 {\zeta'(-1) \over \zeta(-1)}
	+ 2 \log {\Lambda_E \over 4 \pi T} \right)
\right.
\nonumber \\
\left. \;+\; 12 C_A T_F \left( {1 \over \epsilon} + 1 - 2 \log 2
	+ 2 {\zeta'(-1) \over \zeta(-1)}
	+ 2 \log {\Lambda_E \over 4 \pi T} \right)
\right] \;.
\label{deltaf}
\end{eqnarray}
Note that minimal subtraction in the effective theory is not
equivalent to minimal subtraction in the full theory.
In addition to the poles in $\epsilon$ in (\ref{deltaf}),
there are finite terms that depend on the factorization scale $\Lambda_E$.

Matching~(\ref{Fpert}) with~(\ref{Feff}) and using the expression
(\ref{deltaf}), we conclude that $f_E$ to order
$g^4$ is
\begin{eqnarray}
f_E(\Lambda_E) &=& - {\pi^2 d_A \over 9} T^3
\Bigg\{
\left( {1 \over 5} + {7 \over 20} {d_F \over d_A} \right)
\;-\; \left( C_A + {5 \over 2} T_F \right) {g^2(\mu) \over (4 \pi)^2}
\nonumber \\
&+& \Bigg(
C_A^2 \left[ 48 \log{\Lambda_E \over 4 \pi T}
	- {22 \over 3} \log{\mu \over 4 \pi T}
	+ {116\over 5} + 4 \gamma + {148 \over 3} {\zeta'(-1) \over \zeta(-1)}
	- {38 \over 3} {\zeta'(-3) \over \zeta(-3)} \right]
\nonumber \\
&+& C_A T_F
\left[ 48  \log{\Lambda_E \over 4\pi T}
	- {47 \over 3} \log{\mu \over 4\pi T}
	+ {401 \over 60} - {37 \over 5} \log 2
	+ 8 \gamma + {74 \over 3} {\zeta'(-1) \over \zeta(-1)}
	- {1\over 3} {\zeta'(-3) \over \zeta(-3)} \right]
\nonumber \\
&+& T_F^2
\left[ {20 \over 3} \log{\mu \over 4\pi T}
	+ {1 \over 3} - {88 \over 5} \log 2
	+ 4 \gamma + {16 \over 3} {\zeta'(-1) \over \zeta(-1)}
	- {8 \over 3} {\zeta'(-3) \over \zeta(-3)} \right]
\nonumber \\
&+& C_F T_F
\left[ {105 \over 4} - 24 \log 2 \right] \Bigg)
\left( {g^2 \over (4 \pi)^2} \right)^2
\Bigg\} \;,
\label{fE}
\end{eqnarray}
where $g(\mu)$ is the coupling constant in the $\overline{\rm MS}$
renormalization scheme at the scale $\mu$.
We have used (\ref{glam})
to shift the scale of the running coupling constant from $\Lambda_E$ to
an arbitrary renormalization scale $\mu$, and we have identified the
explicit factors of $\Lambda$ that remain
with the factorization scale $\Lambda_E$.

\subsection{Evolution of EQCD coupling constants}

The effective lagrangian (\ref{LEQCD}) for EQCD can be expressed as a sum
over all local operators that respect the symmetries of the theory:
\begin{equation}
f_E(\Lambda_E) \;+\; {\cal L}_{\rm EQCD}
\;=\; \sum_n C_n(\Lambda_E) \; {\cal O}_n ,
\end{equation}
where we have included the unit operator with coefficient
$f_E$ as one of the operators ${\cal O}_n$.
The coefficients $C_n$ are the generalized coupling constants of the effective
theory. Because of ultraviolet divergences, the effective theory
must be regularized with an ultraviolet cutoff $\Lambda_E$.
The ultraviolet divergences in the effective theory include power
ultraviolet divergences proportional to $\Lambda_E^p$, $p=1,2,\ldots$,
and logarithmic divergences proportional to $\log(\Lambda_E)$.
The power divergences are artifacts of the
regularization scheme and have no physical content.
If they are not removed as part of the regularization procedure,
they must be cancelled by power divergences in the coupling constants $C_n$.
In contrast, the logarithmic ultraviolet divergences
are directly related to logarithms of $T$ in the full theory,
and therefore represent real physical effects.
It is convenient to use a regularization procedure for the effective theory
in which power ultraviolet divergences are automatically subtracted,
such as dimensional regularization.  In this case, they need not be cancelled
by power divergences in the coupling constants.
The dimensions of a coupling constant can then only be taken up by powers
of the temperature $T$.  The coupling constant $C_n$ must be proportional
to $T^{3-d_n}$, where $d_n$ is the scaling dimension of the corresponding
operator ${\cal O}_n$.  The dimensionless factor multiplying $T^{3-d_n}$
in the coupling constant $C_n$ can be computed
as a perturbation series in $g^2(T)$, with coefficients that are polynomials
in $\log(T/\Lambda_E)$. The dependence on $\Lambda_E$ is governed by a
``renormalization group equation'' or ``evolution equation'' of the form
\begin{equation}
\Lambda_E {d \ \over d \Lambda_E} C_n(\Lambda_E)
\;=\; \beta_n(C),
\label{dLeff}
\end{equation}
where the beta function $\beta_n$ has a power series expansion in the
coupling constants $C_m$.  These equations follow from the condition
that physical quantities must be independent of the arbitrary scale
$\Lambda_E$.
Since $C_n$ is proportional to $T^{3-d_n}$,
every term in the expansion of its beta function must be proportional
to $T^{3-d_n}$. In particular, a term like $C_{m_1} C_{m_2} \ldots C_{m_k}$
can appear only if the dimensions $d_{m_i}$ of the corresponding operators
${\cal O}_{m_i}$ satisfy
\begin{equation}
\sum_{i=1}^k \left( 3 - d_{m_i} \right) \;=\; 3 - d_n.
\label{cond}
\end{equation}

The condition (\ref{cond}) is very restrictive, particularly if
the effective lagrangian is truncated to the  super-renormalizable
terms that are given explicitly in (\ref{LEQCD}).
It implies that the only terms that can appear in the
beta function for the coefficient $f_E$ of the unit operator
are $g_E^2 m_E^2$, $\lambda_E m_E^2$, and a cubic polynomial in
$g_E^2$ and $\lambda_E$.  Since $m_E^2$, $g_E^2$, and $\lambda_E$
are of order $g^2$, $g^2$, and $g^4$, respectively, the only term of
order $g^4$ is $g_E^2 m_E^2$.  We can determine its coefficient by
calculating the ultraviolet divergences in the strict perturbation
expansion for the free energy in the effective theory.  These divergences
do not appear in (\ref{Feff}), because the ultraviolet poles in
$\epsilon$ have cancelled against infrared poles in $\epsilon$.
We can calculate the ultraviolet divergences by using a different
regularization for infrared divergences.
Alternatively, since we have already calculated $f_E$ explicitly
to order $g^4$, we can simply differentiate (\ref{fE}) and
use the fact that $\Lambda_E (d / d \Lambda_E) f_E$ must be proportional
to $g_E^2 m_E^2$.
Using $g_E^2 = g^2 T$ and the leading order expression for
$m_E^2$ in (\ref{mE}), we find that the evolution equation is
\begin{equation}
\Lambda_E {d \ \over d \Lambda_E} f_E
\;=\; - {d_A C_A \over (4 \pi)^2} g^2_E m^2_E + O(g^6 T^3) \;.
\label{rgfE}
\end{equation}

The beta function for $m_E^2$ must be a quadratic polynomial in $g_E^2$
and $\lambda_E$.  The terms $g_E^4$, $g_E^2 \lambda_E$, and $\lambda_E^2$
are of order $g^4$, $g^6$, and $g^8$, respectively.
The coefficients of these terms can be determined by calculating the
ultraviolet divergent terms in the strict perturbation expansion for
the electric screening mass in the effective theory.  Alternatively,
if $m_E^2$ is known, its beta function can be determined simply by
differentiating.  Since the expression (\ref{mE}) is independent of
$\Lambda_E$, we know that the coefficient of $g_E^4$ in the beta function
vanishes and the leading term must be $g_E^2 \lambda_E$.
Thus the evolution equation for $m_E^2$ is
\begin{equation}
\Lambda_E {d \ \over d \Lambda_E} m_E^2
\;=\; 0 + O(g^6 T^2) \;.
\label{rgmE}
\end{equation}
We have not calculated the coefficient of $g_E^2 \lambda_E$ in this evolution
equation, because it does not affect the free energy until order $g^7$.

The beta functions for $g_E^2$ and $\lambda_E$ vanish to all orders
in the super-renormalizable interactions.  All the nonvanishing
terms in their beta functions involve the coupling constants
of nonrenormalizable interactions, and they are therefore suppressed
by large powers of $g$. The evolution of these paramaters
can probably be ignored for most practical purposes.

The only EQCD parameter whose evolution plays a role in the
free energy to order $g^6$ is $f_E$.  To this order,
the solution to the equation (\ref{rgfE}) is trivial:

\begin{equation}
f_E(\Lambda_E) \;=\; f_E(\Lambda_E')
- {d_A C_A \over (4 \pi)^2} g^2_E m^2_E \log {\Lambda_E \over \Lambda_E'} \;.
\label{rgsol}
\end{equation}

\section{Free Energy to Order $g^5$}

Having calculated the parameters of EQCD to the necessary order in $g^2$,
we now use the effective theory to calculate the free energy to order $g^5$.
The free energy is the sum of the three terms in (\ref{F123}),
which correspond to the momentum scales $T$, $gT$, and $g^2T$, respectively.
The term $f_E T$ is the contribution from the scale $T$.  We have already
calculated $f_E$ to the necessary order and it is given in (\ref{fE}).
The term $f_G T$ is the contribution from the scale $g^2T$, but it does not
contribute until order $g^6$.  The remaining term $f_M T$ is the contribution
>from the scale $gT$.

Through order $g^5$, $f_M$
is proportional to the logarithm of the partition function for EQCD:
\begin{equation}
f_M \;=\; - {\log {\cal Z}_{\rm EQCD} \over V} \;.
\end{equation}
In order to calculate this contribution using perturbation theory,
we must incorporate the terms in the lagrangian that provide
electrostatic screening into the free part of the lagrangian.
The necessary screening effects are provided
by the $A^a_0 A^a_0$ term in the EQCD lagrangian.
Thus we must include the effects of the mass parameter $m_E^2$ to all orders,
while treating all the other coupling constants of EQCD as
perturbation parameters.  The only coupling constant that is required to
obtain the free energy to order $g^5$ is the
gauge coupling constant $g_E$.

The contributions to $\log {\cal Z}_{\rm EQCD}$ of orders
$g^3$, $g^4$, and $g^5$
are given by the sum of the 1-loop, 2-loop, and 3-loop diagrams in
Fig.~6, 7, and 8.  The solid, wavy, dashed lines represent the propagators
of the $A_0$ field, the $A_i$ fields, and the associated ghosts, respectively.
We evaluate these diagrams in Feynman gauge.
They can be expressed in terms of the scalar integrals defined in Appendix~B.
The resulting expression for the logarithm of the partition function is
\begin{eqnarray}
f_M &=& - {d_A \over 2} I'_0
\;+\; d_A C_A g_E^2 \left[ {1 \over 4} I_1^2 + m_E^2 J_1 \right]
\nonumber \\
&& \;+\; d_A C_A^2 g_E^4
\left[ - {1 \over 4} I_1^2 I_2 + 2I_1 J_1 - 2 m_E^2 I_1 J_2
	- m_E^2 I_1 K_2 - {1 \over 4} M_{1,-1}
\right.
\nonumber \\
&& \left.
	- {1-2\epsilon \over 2} M_{0,0} + \epsilon\, M_{-1,1}
	- {1-2\epsilon \over 2} M_{-2,2} + 4 m_E^2 M_{1,0}
	+ 2 m_E^2 M_{0,1}
\right.
\nonumber \\
&& \left.
	- 4 m_E^4 M_{2,0}
	- {3 \over 8} N_{0,0} - {1 \over 2} N_{1,-1}
	- {1 \over 4} N_{2,-2}
	- 2 m_E^2 N_{1,0} - m_E^2 N_{2,-1}
\right.
\nonumber \\
&& \left.
	- m_E^4 N_{1,1} - m_E^4 N_{2,0} - {1 \over 4} L_{1,-1} \right]
\;+\; \delta f_E \;,
\label{freeint}
\end{eqnarray}
where $\delta f_E$ is the counterterm associated with the unit operator
of the EQCD lagrangian.
The integrals $I_n$, $J_n$, $K_n$, $L_{m,n}$, $M_{m,n}$, and $N_{m,n}$
can be calculated analytically
using methods developed by Broadhurst \cite{broadhurst}
and they are given in Appendix~B. Adding them up,
we obtain
\begin{eqnarray}
f_M
\;=\;  \;-\; {d_A \over 3 (4 \pi)} m_E^3
\;+\; {d_A C_A \over 4 (4 \pi)^2}
\left( {1 \over \epsilon} + 4 \log {\Lambda \over 2 m_E} + 3 \right)
	g_E^2 m_E^2
\nonumber \\
\;+\; {d_A C_A^2 \over (4 \pi)^3}
\left( {89 \over 24} - {11 \over 6} \log 2
	+ {1 \over 6} \pi^2 \right) g_E^4 m_E
\;+\; \delta f_E \;,
\label{logZeff3}
\end{eqnarray}
where $\Lambda$ is the scale of dimensional regularization.
It can be identified with the ultraviolet cutoff $\Lambda_E$ of EQCD.
The ultraviolet pole in $\epsilon$ in the term proportional to $g^2_E m^2_E$
in (\ref{logZeff3}) is cancelled by the counterterm
$\delta f_E$, which is given in (\ref{deltafE}).
Our final result is therefore
\begin{eqnarray}
f_M(\Lambda_E) \;=\; - {d_A \over 3(4 \pi)} m_E^3 \;
\Bigg\{ 1
\;+\; \left[-3 \log {\Lambda_E \over 2 m_E} - {9 \over 4} \right]
	{C_A g_E^2 \over 4 \pi m_E}
\nonumber \\
\;+\; \left[ - {89 \over 8} + {11 \over 2} \log 2  - {1\over 2} \pi^2 \right]
	\left({C_A g_E^2 \over 4 \pi m_E}\right)^2
\Bigg\}\,.
\label{fM}
\end{eqnarray}

The coefficient $f_M$ in (\ref{fM}) can be expanded in powers of $g$
by setting $g_E^2 = g^2T$ and by substituting the expression (\ref{mE})
for $m_E^2$.  The complete free energy to order $g^5$ is then
$F = (f_E + f_M) T$.
Note that the dependence on the arbitrary factorization scale
$\Lambda_E$ cancels between $f_E$ and $f_M$, up to corrections
that are higher order in $g$, leaving a logarithm of $T/m_E$.
This $g^4 \log(g)$ term is associated with the renormalization of
$f_E$, and its coefficient can be determined from the evolution
equation (\ref{rgfE}).  There is no $g^5 \log(g)$ term in the
perturbation expansion for $F$, and this is a consequence of
the vanishing of the order-$g^4$ term in the beta function for $m_E^2$.

\section{Outline of Calculation to Order $g^6$}

The calculation of the free energy to order $g^5$, which was presented
in the previous section, was greatly streamlined by using effective
field theory to unravel the effects of the momentum scales $T$ and $gT$.
The same result has also been obtained by Kastening and Zhai using other
methods \cite{kastening-zhai}. However
the advantages of the effective field theory approach become
more and more apparent as we go to higher order in $g$.
In this section, we demonstrate the power of this method by
outlining the calculation of the free energy to order $g^6$.
In this case there are contributions from all three momentum scales $T$,
$gT$, and $g^2T$.

\subsection{Contribution from the scale $g^2T$}

We first discuss the contribution to the free energy from the scale
$g^2T$, which is given by the term $f_G T$ in (\ref{F123}).  This term is
proportional to the logarithm of the partition function (\ref{ZMQCD}) of
MQCD.  Treating the correction term $\delta {\cal L}_{\rm MQCD}$ in the
MQCD lagrangian as a perturbation, the partition function can be written
\begin{equation}
{\cal Z}_{\rm MQCD} \;=\; \int^{(\Lambda_M)} {\cal D} A^a_i({\bf x})
\exp \left( - \int d^3x \; G^2/4 \right)
\left\{ 1 \;-\; \int d^3x \; \delta {\cal L}_{\rm MQCD} \;+\; \ldots \right\}
\;,
\label{ZMQCD1}
\end{equation}
where $G^2 \equiv G^a_{ij} G^a_{ij}$.
Taking the logarithm of both sides, we obtain
\begin{equation}
f_G \;=\;
- { \log {\cal Z}_{\rm MQCD}^{(0)} \over V }
	\;+\; \langle \delta {\cal L}_{\rm MQCD} \rangle_0
	\;+\; \ldots \;,
\label{fG01}
\end{equation}
where ${\cal Z}_{\rm MQCD}^{(0)}$ is the partition function for the
minimal gauge theory with action $\int d^3x \, G^2/4$.
The subscript 0 on the expectation value
$\langle \delta {\cal L}_{\rm MQCD} \rangle_0$ is a reminder that it is
to be calculated using the minimal gauge theory action.

For the moment, let us consider only the first term in (\ref{fG01}).
The  partition function ${\cal Z}_{\rm MQCD}^{(0)}$
is that of the minimal gauge theory
in 3 dimensions.  This is a super-renormalizable theory and its ultraviolet
divergences have a very simple structure.  By naive power-counting,
ultraviolet divergences in $\log {\cal Z}_{\rm MQCD}^{(0)}$ can arise
only from vacuum diagrams with 1, 2, 3, or 4 loops or from propagator
corrections with 1 or 2 loops.  Ward identities guarantee
that the propagator corrections are actually finite.
This is related to the fact that the only gauge invariant operator
with dimension lower than $G^2$ is the unit operator.
Thus the only ultraviolet divergences are in the vacuum diagrams.
The 1-loop diagrams give a cubic divergence.  The 2-loop diagrams
give a quadratic divergence proportional to $g_M^2$.  The 3-loop
diagrams give a linear divergence proportional to $g_M^4$.  Finally,
the 4-loop diagrams give a logarithmic divergence proportional to
$g_M^6$.  After subtraction of the power divergences, we can use
dimensional analysis to determine the form of
$\log {\cal Z}_{\rm MQCD}^{(0)}$.  Aside from the logarithmic dependence
on the ultraviolet cutoff $\Lambda_M$, the only scale in the problem is $g_M$.
By dimensional analysis, $\log {\cal Z}_{\rm MQCD}^{(0)}$
must be proportional to $g_M^6$.  Thus it must have the form
\begin{equation}
- { \log {\cal Z}_{\rm MQCD}^{(0)} \over V }
\;=\; \left( a + b \log {\Lambda_M \over g_M^2} \right)
	 g_M^6  \;,
\label{ZMQCD0}
\end{equation}
where $a$ and $b$ are pure numbers.  The coefficient $b$ can be
determined by calculating the logarithmic ultraviolet divergence in the
4-loop vacuum diagrams for MQCD. The coefficient $a$ can only be calculated
using nonperturbative methods.  It can for example be extracted from
measurements of the expectation value $\langle G^2 \rangle_0$ using
lattice simulations of the pure gauge theory.
A convenient expression for $\langle G^2 \rangle_0$ can be obtained
by taking the logarithm of both sides of (\ref{ZMQCD0}) and
differentiating with respect to $g_M^2$.  It is useful to first
rescale the field $A_i$ in the functional integral for
${\cal Z}_{\rm MQCD}^{(0)}$, so that the coupling constant appears
only in the coefficient $1/g_M^2$ of the action.
After subtracting the power ultraviolet divergences,
we obtain the expression
\begin{equation}
\langle G^2 \rangle_0 \;=\;
-4 \left( 3 a - b + 3 b \log {\Lambda_M \over g_M^2} \right) g_M^6 \;.
\label{G20}
\end{equation}
The subscript 0 on the expectation value $\langle G^2 \rangle_0$ is a
reminder that it is to be calculated using the minimal gauge theory action
$\int d^3x \, G^2/4$ rather that the full action of MQCD.
The expectation value $\langle G^2 \rangle_0$ can be measured on the lattice
using Monte Carlo simulations of the minimal gauge theory.  Once
$\langle G^2 \rangle_0$ has been measured and the
coefficient $b$ has been calculated, we can determine $a$  using the
formula (\ref{G20}).

We now verify that the correction term in (\ref{ZMQCD1}) from higher
dimension operators in the MQCD lagrangian can
indeed be treated as a small perturbation.
The lowest dimension operators in $\delta {\cal L}_{\rm MQCD}$ are
$G^3 \equiv f^{abc} G^a_{ij} G^b_{jk} G^c_{ki}$,
whose coefficient is proportional to $g^3/T^{3/2}$,
and $(DG)^2 \equiv (D_i G_{ik})^a (D_j G_{jk})^a$,
whose coefficient is proportional to $g^2/T^2$.
Their coefficients have been calculated to leading order in $g$ by
Chapman for the case of a pure gauge theory \cite{chapman}.
After subtraction of power ultraviolet divergences, the only
scale in the problem is $g_M^2$. Therefore,
by dimensional analysis, $\langle G^3 \rangle_0$ must be proportional to
$(g_M^2)^{9/2}$.  Using $g_M^2 \approx g^2 T$ and taking into account
the coefficient which is proportional to $g^3/T^{3/2}$,
we find that the contribution to $f_G$ from $\langle G^3 \rangle_0$
is of order $g^{12}T^3$.  Using a similar analysis, we find that the
contribution from $\langle (DG)^2 \rangle_0$ is also of order $g^{12} T^3$.
Thus the effects of higher dimension operators in the MQCD lagrangian are
indeed suppressed by powers of the coupling constant $g$.

We have found that the contribution to the free energy from the scale
$g^2T$ can be written
\begin{equation}
f_G T \;=\;
\left( a + b \log{\Lambda_M \over g_M^2} \right) g_M^6 T
	\;+\; O(g^{12} T^4) \;,
\label{fG}
\end{equation}
Remarkably, the only nonperturbative calculation that is required
to determine the free energy up to order $g^{12}$ is that of the single
pure number $a$.  We also require the coupling constant $g_M$, which
can be calculated by matching perturbative calculations in EQCD and MQCD.
To calculate the free energy to order $g^6$, we only need $g_M$ to leading
order in $g$.  At this order, it is given simply by $g_M^2 = g^2 T$.
In summary, in order to obtain the contribution to the free energy from
the scale $g^2T$ to order $g^6$,  all that is required are the two pure
numbers $a$ and $b$ in (\ref{fG}).  The number $b$ can be calculated by
evaluating 4-loop diagrams in MQCD.  In Ref. \cite{solution}, it was
assumed incorrectly that this number vanishes.  The number $a$ can be
calculated using lattice simulations of the pure gauge theory in 3
dimensions.

\subsection{Contribution from the scale $gT$}

The contribution to the free energy from the scale $g T$ is given by the
term $f_M T$ in (\ref{F123}).  The coefficient $f_M$ can be determined by
calculating the logarithm of the EQCD partition
function in both EQCD and MQCD and matching the expressions.
If we use dimensional regularization
to cut off both infrared and ultraviolet divergences, all the loop diagrams
in MQCD vanish.  The expression for the logarithm of the partition function
then is simply
\begin{equation}
- {\log {\cal Z}_{\rm EQCD} \over V} \;=\;
f_M + \delta f_M \;,
\label{ZEQCDM}
\end{equation}
where $\delta f_M$ is a counterterm that cancels ultraviolet divergences
in MQCD that are proportional to the
unit operator.  To order $g^6$, this counterterm is simply
\begin{equation}
\delta f_M \;=\; {b \over 2 \epsilon} g_M^6 \;,
\label{deltafM}
\end{equation}
where $b$ is the same coefficient that appears in (\ref{fG}).

To determine $f_M$, we must match the expression (\ref{ZEQCDM}) with the
corresponding expression in EQCD, which is obtained by calculating the
sum of vacuum diagrams using dimensional regularization.  The resulting
expression for $\log {\cal Z}_{\rm EQCD}$ is a sum of
polynomials in the EQCD coupling constants, such as
$g_E^2$ and $\lambda_E$, multiplied by whatever powers of $m_E$ are required
by dimensional analysis.  There are three such terms that contribute
to the free energy at order $g^6$.
The first term is $g_E^2 m_E^2$, whose coefficient has already been
calculated in (\ref{fM}).  It contributes through the next-to-leading order
term in $m_E^2$, which is given in (\ref{mE}), and through the
next-to-leading order term in $g_E^2$, which has not yet been calculated.
The second term which contributes at order $g^6$ is proportional to $g_E^6$.
Its coefficient is determined by calculating all 4-loop vacuum diagrams
that involve only the gauge coupling constant $g_E$.
This term will have a pole in $\epsilon$ that matches that from the
counterterm  (\ref{deltafM}).
The third term that contributes to $f_M$ at order $g^6$
is proportional to $\lambda_E m_E^2$.  Its coefficient
is given by the single 2-loop vacuum diagram that involves the $A_0^4$
coupling constant $\lambda_E$.
This coupling constant is only required to leading order in $g$
and has already been calculated by Nadkarni and by Landsman
\cite{nadkarni-2,landsman}.

In summary, there are three coefficients that must be calculated in order
to obtain the contribution of order $g^6$ to the free energy from the
scale $gT$.  We need the coefficients of $g_E^6$ and of $\lambda_E m_E^2$
in the expression for $f_M$.  These can be  obtained by
perturbative calculations in EQCD.  We also need the coefficient of $g^4$
in the expression for the EQCD parameter $g_E^2$.  This
requires a perturbative calculation in full QCD.

\subsection{Contribution from the Scale $T$}

The contribution to the free energy from the scale $T$ is given by the
term $f_E T$ in (\ref{F123}).  The term $f_E$ is obtained by matching
the strict perturbation expansions for the free energy in full QCD
and in EQCD.   In full QCD, the contribution
of order $g^6$ is the sum of all 4-loop vacuum diagrams.
If we use dimensional regularization to cut off
both infrared and ultraviolet divergences, then the corresponding
expression in EQCD is simply
$F = (f_E + \delta f_E) T$.  The counterterm $\delta f_E$ includes the term
proportional to $g_E^2 m_E^2/\epsilon$ given in (\ref{deltafE})
and also a term proportional to $\lambda_E m_E^2/\epsilon$.
Since the counterterm is proportional to
$1/\epsilon$, we need not only the value of the coupling constant
$\lambda_E$ at $\epsilon =0$ but also the terms linear in $\epsilon$.
Similarly, we need the term of order $\epsilon$ in the order-$g^4$
correction to $g_E^2$.

In summary, there are several calculations that must be carried out
in order to obtain the term of order $g^6$ in $f_E$.  We need to
calculate the 4-loop vacuum diagrams in full QCD.  We also need to
calculate the terms of order $\epsilon g^4$ in the EQCD parameters
$g_E^2$ and $\lambda_E$.

\section{Convergence of Perturbation Theory}

We have calculated the free energy as a perturbation expansion
in powers of $g$ to order $g^5$.  In this section, we examine
the convergence of that perturbation expansion.
For simplicity, we  focus on the case of QCD with $n_f$ flavors of quarks.

The effects of the momentum scale $T$ enter into the free energy only
through the coefficient $f_E$ and the parameters in the EQCD lagrangian.
The term $f_E$ is given in (\ref{fE}):
\begin{eqnarray}
f_E(\Lambda_E) &=&
- {8 \pi^2 \over 45} T^3
\Bigg\{ 1 + {\textstyle{21 \over 32}} n_f
\;-\; {15 \over 4} \left(1 + {\textstyle{5 \over 12}} n_f \right)
	{\alpha_s(\mu) \over \pi}
\nonumber \\
&&
\;+\; \left[ 244.9 - 17.24 n_f - 0.415 n_f^2
\;-\; {165 \over 8} \left( 1 + {\textstyle{5 \over 12}} n_f \right)
	\left( 1 - {\textstyle{2 \over 33}} n_f \right)
	\log {\mu \over 2 \pi T}
\right.
\nonumber \\
&& \left.
\;-\; 135 \left(1 + {\textstyle{1 \over 6}} n_f \right)
		\log {\Lambda_E \over 2 \pi T} \right]
	\left( {\alpha_s \over \pi} \right)^2
\;+\; O(\alpha_s^3) \Bigg\} \;.
\label{fEnum}
\end{eqnarray}
The other parameters in the EQCD lagrangian that enter into the
calculation of the free energy to order $g^5$ are $m_E$ and $g_E$,
which are given by (\ref{mE}) and (\ref{gE}), respectively:
\begin{eqnarray}
m_E^2 &=&
4 \pi \; \alpha_s(\mu) \; T^2
\Bigg\{ 1 + \textstyle{1 \over 6} n_f
\;+\; \Bigg[ 0.612 - 0.488 n_f - 0.0428 n_f^2
\nonumber \\
&& \;+\; {11 \over 2} \left(1 + {\textstyle{1 \over 6}} n_f \right)
	\left(1 - {\textstyle{2 \over 33}} n_f \right)
	\log {\mu \over 2 \pi T} \Bigg]
	{\alpha_s \over \pi}
\;+\; O(\alpha_s^2) \Bigg\} \;,
\label{mEnum}
\\
g_E^2 &=&
4 \pi \; \alpha_s \; T \left[1 \;+\; O(\alpha_s) \right] \;.
\label{gEnum}
\end{eqnarray}

We have calculated two terms in the perturbation series for $m_E^2$
and three terms in the series for $f_E$.  We can use these results to
study the convergence of perturbation theory for the parameters
of EQCD.  We consider the case of $n_f=3$ flavors of quarks, although our
conclusions will not depend sensitively on $n_f$.  The question of
the convergence is complicated by the presence of the
arbitrary renormalization and factorization scales $\mu$ and $\Lambda_E$.
The next-to-leading-order (NLO) correction to $f_E$ is independent
of $\mu$ and $\Lambda_E$, and is small compared to the leading-order (LO)
term provided that $\alpha_s(\mu) \ll 1.1$.
The NLO correction to $m_E^2$ and the next-to-next-to-leading-order (NNLO)
correction to $f_E$ both depend on the renormalization scale $\mu$.
One scale-setting scheme that is physically well-motivated is
the BLM prescription \cite{BLM}, in which $\mu$ is adjusted to cancel
the highest power of $n_f$ in the correction term.
This prescription gives $\mu = 0.93 \, \pi T$ when applied to $m_E^2$
and $\mu = 4.4 \, \pi T$ when applied to $f_E$.  These values
differ only by about a factor of 2 from $2 \pi T$, which is the lowest
Matsubara frequency for gluons.  Below, we will consider the three values
$\mu = \pi T$, $2 \pi T$, and $4 \pi T$.
For the NLO correction to $m_E^2$ to be much smaller than the LO term,
we must have $\alpha_s(\mu) \ll$ 0.8, 3.8, and 1.4 if
$\mu = \pi T$, $2 \pi T$, and $4 \pi T$, respectively.
Based on these results,
we conclude that the perturbation series for the parameters of
EQCD are well-behaved provided that $\alpha_s(2 \pi T) \ll 1$.

The NNLO correction for $f_E$ depends not only on $\mu$, but also on the
factorization scale $\Lambda_E$.  Because the coefficient of
$\log(\Lambda_E/2 \pi T)$
in (\ref{fE}) is so much larger than that of $\log(\mu/2 \pi T)$,
the NNLO correction for $f_E$
is much more sensitive to $\Lambda_E$ than to $\mu$.
It is useful intuitively to think of the
infrared cutoff $\Lambda_E$ as being much smaller than the
ultraviolet cutoff $\mu$.  However, these scales can be identified
with momentum cutoffs only up to multiplicative constants that may be
different for $\mu$ and $\Lambda_E$.  Both parameters are introduced
through dimensional regularization, but $\mu$ arises from ultraviolet
divergences of 4-dimensional integrals, while $\Lambda_E$ arises
>from infrared divergences of 3-dimensional integrals.
We might be tempted to set $\Lambda_E = \mu$, but then the
NNLO coefficient in $f_E$ is large.  For the choice
$\mu = 2 \pi T$, the correction to the LO term is a multiplicative factor
$1 - 0.9 \alpha_s + 6.5 \alpha_s^2$.  The NNLO correction can be made small
by adjusting $\Lambda_E$.
It vanishes for $\Lambda_E = 5.8 \, \pi T$, $5.1 \,\pi T$,
and $4.5 \, \pi T$ if $\mu = \pi T$, $2 \pi T$, and $4 \pi T$, respectively.
We conclude that the perturbation series for $f_E$ is well-behaved
if the factorization scale $\Lambda_E$ is chosen to be approximately
$5 \pi T$.  Whether this choice is reasonable
can only be determined by calculating other EQCD parameters to higher order
to see if the same choice leads to well-behaved perturbation series.

The choice of $\Lambda_E$ that makes the perturbation series for the
EQCD parameters well-behaved may be much larger than the largest mass
scale $m_E$ of EQCD.  Perturbative corrections in EQCD will then
include large logarithms of $\Lambda_E/m_E$.  This problem can be avoided
by using renormalization group equations to evolve the parameters
of EQCD from the initial scale $\Lambda_E$ down to some scale $\Lambda_E'$
of order $m_E$.
The solution to  the renormalization group equation for $f_E$ is given
in (\ref{rgsol}).
The evolution of $g_E^2$ and $m_E^2$ occurs only at
higher order in the coupling constant and therefore can be ignored.

We have carried out only one perturbative calculation in EQCD.
This is the term $f_M$, which gives the contribution to the
free energy from the scale $gT$.
This term is given in (\ref{fM}):
\begin{equation}
f_M(\Lambda_E) \;=\; - {2 \over 3 \pi} m_E^3
\left[ 1 \;-\; \left( 0.256 - {9 \over 4} \log{\Lambda_E \over m_E} \right)
	{g_E^2 \over 2 \pi m_E}
	\;-\; 27.6 \left( {g_E^2 \over 2 \pi m_E} \right)^2
	\;+\; O(g^3) \right] \;.
\label{fMnum}
\end{equation}
We now consider the convergence of the perturbation series (\ref{fM})
for $f_M$.  The size of the NLO correction depends on the choice of the
factorization scale $\Lambda_E$.  It is small if $\Lambda_E$ is chosen to be
approximately $m_E$.  The NNLO correction in (\ref{fM}) is independent
of any arbitrary scales.  If $n_f=3$, it is small compared to the leading order
term only if $\alpha_s \ll 0.17$.  Thus the perturbation series
for $f_M$ is well-behaved only for values of
$\alpha_s(2 \pi T)$ that are much smaller than those required
for the parameters of EQCD to have well-behaved perturbation series.

Inserting (\ref{mEnum}) and (\ref{gEnum}) into (\ref{fMnum}),
expanding in powers of $g$, and adding (\ref{fEnum}), we get the
expansion for the free energy in powers of $\sqrt{\alpha_s}$:
\begin{eqnarray}
F \;=\; - {8 \pi^2 \over 45} T^4\;
\left[ F_0
\;+\; F_2  {\alpha_s(\mu) \over \pi}
\;+\; F_3  \left( {\alpha_s(\mu) \over \pi} \right)^{3/2}
\;+\; F_4  \left( {\alpha_s \over \pi} \right)^2
\right.
\nonumber \\
\left.
\;+\; F_5  \left( {\alpha_s \over \pi} \right)^{5/2}
\;+\; O(\alpha_s^3 \log \alpha_s) \right] \;.
\label{freeg}
\end{eqnarray}
The coefficients in this expansion are
\begin{eqnarray}
F_0 &=& 1 + \textstyle{21 \over 32} n_f \;,
\\
F_2 &=& - {15 \over 4} \left( 1 + \textstyle{5 \over 12} n_f \right)\;,
\\
F_3 &=& 30 \left( 1 + \textstyle{1 \over 6} n_f \right)^{3/2} \;,
\\
F_4 &=&  237.2 + 15.97 n_f - 0.413 n_f^2
+ { 135 \over 2} \left( 1 + \textstyle{1 \over 6} n_f \right)
	\log \left[ {\alpha_s \over \pi}
		\left(1 + \textstyle{n_f \over 6} \right) \right]
\nonumber \\
&& \;-\; { 165 \over 8}
\left( 1 + \textstyle{5 \over 12} n_f \right)
\left( 1 - \textstyle{2 \over 33} n_f \right)
	\log {\mu \over 2 \pi T} \;,
\\
F_5 &=& \left( 1 + \textstyle{1 \over 6} n_f \right)^{1/2}
\Bigg[ -799.2 - 21.96 n_f - 1.926 n_f^2
\nonumber \\
&& \;+\; {495 \over 2} \left( 1 + \textstyle{1 \over 6} n_f \right)
	\left( 1 - \textstyle{2 \over 33} n_f \right)
		\log {\mu \over 2 \pi T} \Bigg] \;.
\end{eqnarray}
The coefficient $F_2$ was first given by Shuryak \cite{shuryak}.
The coefficient of $F_3$ was first calculated correctly
by Kapusta \cite{kapusta}.
The coefficient $F_4$ was calculated in 1994
by Arnold and Zhai \cite{arnold-zhai}.
The coefficient $F_5$ has also been calculated independently by
Kastening and Zhai \cite{kastening-zhai}.

We now ask how small $\alpha_s$ must be in order for the expansion
(\ref{freeg}) to be well-behaved.
For simplicity, we consider the case $n_f=3$,
although our conclusions are not sensitive to $n_f$.
If we choose the renormalization scale $\mu = 2 \pi T$ motivated
by the BLM criterion \cite{BLM},
the correction to the LO result is a multiplicative factor
$1 - 0.9 \alpha_s + 3.3 \alpha_s^{3/2}
	+ (7.1 + 3.5 \log \alpha_s) \alpha_s^2 - 20.8 \alpha_s^{5/2}$.
The $\alpha_s^{5/2}$ term is the largest correction unless
$\alpha_s(2 \pi T) < 0.12$.
We can make the $\alpha_s^{5/2}$ term small only by choosing the
renormalization scale to be near the value $\mu = 36.5 \pi T$
for which $F_5$ vanishes.  This ridiculously large of $\mu$ arises because
the scale $\mu$ has been adjusted to cancel the large $g^5$ correction
to $f_M$ in (\ref{fM}).  This contribution arises from the momentum
scale $gT$ and has nothing to do with renormalization of $\alpha_s$.
We conclude that the expansion (\ref{freeg}) for $F$ in powers of
$\sqrt{\alpha_s}$ is well-behaved only if $\alpha_s(2 \pi T) \ll 1/10$.
This is an order of magnitude smaller than the value required for the
EQCD parameters to be well-behaved.  Our previous analysis indicates
that this slow convergence  of the expansion for $F$ in powers
of $\sqrt{\alpha_s}$ can be attributed to the slow convergence of
perturbation theory at the scale $gT$.

\section{Discussion}

In this paper, we have used effective-field-theory methods to
unravel the contributions to the free energy of high temperature
QCD from the scales
$T$, $gT$, and $g^2T$.  We calculated the free energy explicitly
to order $g^5$.  The calculation was significantly streamlined
by using effective-field-theory methods to reduce
every step of the calculation to one that involves only a single
momentum scale.
We also outlined the calculations that would be necessary to
obtain the free energy to order $g^6$.  It is only at this order
that the full power of the effective-field-theory approach
becomes evident.

The effective-field-theory approach provides an understanding
of the logarithms of the coupling constant that arise in
perturbation expansions in thermal field theory.
These logarithms are associated with the renormalization of
the parameters of effective field theories. The resulting
evolution equations can be used to sum up leading logarithms
of the coupling constant of the form $g^{m+2n} \log^n(g)$ to all orders
in $n$ \cite{eft}.
To the accuracy required for the calculation of the free energy
to order $g^5$ in QCD, this resummation is trivial. The only terms of the
form $g^{m+2n} \log^n(g)$ with $m+2n \le 5$ are a $g^4 \log(g)$ term
associated with renormalization of the coefficient $f_E$.  The fact that the
solution (\ref{rgsol}) to the evolution equation for $f_E$ is trivial
indicates that there are no higher order
terms of the form $g^{2+2n} \log^n(g)$ that are related to the
$g^4 \log(g)$ term through the renomalization group.
There are also no terms of the form $g^{3+2n} \log^n(g)$ in the free energy.
This is a consequence of the vanishing of the  $g_E^4$ term
in the beta function for $m_E^2$.  In the seemingly simpler problem of a
massless scalar field with a $\phi^4$ interaction,
the evolution equations play a more important role  \cite{eft}.
There are terms in the free energy
of the form $g^{3+2n} \log^n(g)$ that can be summed up to all orders
with the help of the renormalization group.  The relative simplicity
of the QCD case comes from the fact that the term $g_E^4$ in the
beta function for $m_E^2$ has a vanishing coefficient.  We know
of no deep reason for this coefficient to vanish.

Our explicit calculations allow us to study the convergence of the
perturbation expansion for thermal QCD.  They suggest that
perturbation theory at the scale $gT$ requires a much smaller value
of the coupling constant than perturbation theory at the scale $T$.
At the scale $T$, perturbation corrections can be small only if
$\alpha_s(2 \pi T) \ll 1$.
Of course, even if this condition is satisfied,
the perturbation expansion may break down anyway,
but this is certainly a necessary condition.
At the scale $gT$, perturbation corrections can be small only if
$\alpha_s(2 \pi T) \ll 1/10$.  Thus, in order to achieve a given
relative accuracy, the coupling constant $\alpha_s(2 \pi T)$ must be
an order of magnitude smaller for perturbation theory at the scale $gT$
compared to perturbation theory at the scale $T$.
This has important implications for calculations in thermal QCD.
At extremely high temperatures,
the asymptotic freedom of QCD guarantees that the running coupling
constant $\alpha_s(2 \pi T)$ is sufficiently small that perturbation
theory will provide an accurate treatment of the effects of the scale
$gT$ as well as those of the scale $T$. Nonperturbative
methods, such as lattice simulations of MQCD, are necessary only to
calculate the effects of the scale $g^2T$.
Of course, one can always treat the entire problem nonperturbatively by
carrying out lattice simulations of full thermal QCD.  However it is probably
more efficient to integrate out the scales $T$ and $gT$ using perturbative
methods, and to reserve the nonperturbative methods only for the scale $g^2T$
where they are essential.
As the temperature is decreased, the running coupling constant increases
and perturbation theory becomes less accurate.
At sufficiently low temperatures, perturbation theory
breaks down completely, and the entire problem must be treated
nonperturbatively.  This is certainly the case when the temperature
is close to the critical temperature for the phase transition  from the
quark-gluon plasma to a  hadron gas.

Our calculations suggest, however, that there
is a range of temperatures in which perturbation theory
at the scale $gT$ has broken down, but perturbation theory at the scale $T$
is reasonably accurate. In this case, one can still use perturbation
theory at the scale $T$ to calculate the parameters in the EQCD
lagrangian.  Our calculations of the coefficients $f_E$ and $m_E^2$
to order $g^4$ are therefore still useful.  However, nonperturbative methods,
such as  lattice simulations of EQCD, are required to
calculate the effects of the smaller momentum scales $gT$ and $g^2T$.
While one could simply treat the entire problem nonperturbatively using
lattice simulations of full QCD, the effective-field-theory approach provides
a dramatic savings in resources for numerical computation.  The savings come
>from two sources. One is the reduction of the problem from a 4-dimensional
field theory to a 3-dimensional field theory.  The other source of savings
is that quarks are integrated out of the theory, which
reduces it to a purely bosonic problem.

We now consider briefly the implications for the study
of the quark-gluon plasma in heavy-ion collisions.
The critical temperature $T_c$ for formation
of a quark-gluon plasma is approximately 200 MeV.  It may be possible
in heavy-ion collisions to produce a quark-gluon plasma with
temperatures several times $T_c$.  At $T = 350 \; {\rm MeV}$,
$\alpha_s(2 \pi T) \approx 0.3$, which is
small enough that perturbation theory
may be reasonably convergent at the scale $T$, but it is certainly
not convergent at the scale $gT$.  We conclude that at the temperatures
achievable in heavy-ion collisions, perturbative QCD may be accurate
when applied to quantities that involve the scale $T$ only.
However nonperturbative methods are required to accurately calculate
quantities that involve the scales $gT$ and $g^2T$.
The most effective strategy for calculating the properties of a
quark-gluon plasma at such temperatures will probably involve
a combination of perturbative and nonperturbative methods.
The effective-field-theory approach developed in this paper provides
a systematic method for unraveling the momentum scales in the plasma
and for combining perturbative and nonperturbative methods in a consistent way.
This approach applies strictly only to static properties
and to the case of zero baryon density.
The extension to dynamical properties and to the case
of nonzero baryon density remains a challenging problem.

\bigskip

\section*{Acknowledgements}

This work was supported in part by the U.~S. Department of Energy,
Division of High Energy Physics, under Grant DE-FG02-91-ER40684,
and by the Ministerio de Educaci\'on y Ciencia of Spain.

\appendix\bigskip\renewcommand{\theequation}{\thesection.\arabic{equation}}
\section{Sum-integrals in the Full Theory}
\setcounter{equation}{0}\label{ap:a}

In the imaginary-time formalism for thermal field theory, the 4-momentum
$P=(p_0,{\rm\bf p})$ is euclidean with $P^2 = p_0^2+{\rm\bf p}^2$. The
euclidean energy $p_0$ has discrete values: $p_0 = 2 n \pi T$ for bosons
and $p_0 = (2n+1) \pi T$ for fermions, where $n$ is
an integer. Loop diagrams involve sums over $p_0$ and integrals over
{\bf p}. It is convenient to use dimensional regularization
to regularize both ultraviolet and infrared divergences.
We introduce a concise notation for these
regularized sum-integrals:
\begin{eqnarray}
\hbox{$\sum$}\!\!\!\!\!\!\int_P &\equiv&
\left({e^\gamma \Lambda^2 \over 4 \pi}\right)^\epsilon\;
T \sum_{p_0 = 2n \pi T} \:\int {d^{3-2\epsilon}p \over (2
\pi)^{3-2\epsilon}}\,,
\\
\hbox{$\sum$}\!\!\!\!\!\!\int_{\{P\}} &\equiv&
\left({e^\gamma \Lambda^2 \over 4 \pi}\right)^\epsilon\;
T \sum_{p_0 = (2n+1) \pi T} \:\int {d^{3-2\epsilon}p \over (2
\pi)^{3-2\epsilon}}\,,
\end{eqnarray}
where $3-2\epsilon$ is the dimension of space and $\Lambda$ is an arbitrary
momentum scale.
The factor $(e^\gamma/4\pi)^\epsilon$
is introduced so that, after minimal subtraction of the poles in $\epsilon$
due to ultraviolet divergences, $\Lambda$ coincides with the renormalization
scale in the $\overline{\rm MS}$ renormalization scheme.
Below, we collect together all the sum-integrals that are required
to calculate the coefficient $f_E$ to next-to-next-to-leading order
in $g^2$ and the coefficient $m_E^2$ to next-to-leading order in $g^2$.

The one-loop bosonic sum-integrals that arise in the calculation
have the following forms:
\begin{eqnarray}
{\cal I}_n &\equiv& \hbox{$\sum$}\!\!\!\!\!\! \int_P {1 \over (P^2)^n} ,
\\
{\cal J}_n &\equiv& \hbox{$\sum$}\!\!\!\!\!\! \int_P {p_0^2 \over (P^2)^{n+1}}
,
\\
{\cal K}_n &\equiv& \hbox{$\sum$}\!\!\!\!\!\! \int_P {p_0^4 \over (P^2)^{n+2}}
{}.
\end{eqnarray}
The specific sum-integrals that are needed are
\begin{eqnarray}
{\cal I}'_0 &=& {\pi^2 \over 45} T^4 \left[ 1 + O(\epsilon) \right]\,,
\label{I0prime}
\\
{\cal I}_1 &=& {1 \over 12} T^2
\left( {\Lambda \over 4 \pi T} \right)^{2 \epsilon}
\left[1 +  \left( 2 + 2 {\zeta'(-1) \over \zeta(-1)} \right) \epsilon
	+ O(\epsilon^2) \right]\,,
\\
{\cal J}_1 &=& - {1 \over 24} T^2
\left( {\Lambda \over 4 \pi T} \right)^{2 \epsilon}
\left[ 1 + 2 {\zeta'(-1) \over \zeta(-1)} \epsilon + O(\epsilon^2) \right] \,,
\\
{\cal I}_2 &=& {1 \over (4\pi)^2}
\left( {\Lambda \over 4 \pi T} \right)^{2 \epsilon}
\left[ {1 \over \epsilon} + 2 \gamma + O(\epsilon) \right]\,,
\label{a3}
\\
{\cal J}_2 &=& {1 \over 4 (4 \pi)^2}
\left( {\Lambda \over 4 \pi T} \right)^{2 \epsilon}
\left[ {1 \over \epsilon} + 2 + 2 \gamma + O(\epsilon) \right]\,,
\\
{\cal K}_2 &=& {1 \over 8 (4 \pi)^2}
\left( {\Lambda \over 4 \pi T} \right)^{2 \epsilon}
\left[ {1 \over \epsilon} + {8 \over 3} + 2 \gamma + O(\epsilon) \right]\,,
\end{eqnarray}
where $\gamma$ is Euler's constant and $\zeta(z)$ is Riemann's
zeta function.
In  (\ref{I0prime}), ${\cal I}'_0$ denotes the derivative of
${\cal I}_n$ with respect to $n$ evaluated at $n=0$.
The one-loop fermionic sum-integrals have the following forms:
\begin{eqnarray}
\widetilde{\cal I}_n &\equiv&
	\hbox{$\sum$}\!\!\!\!\!\! \int_{\{P\}} {1 \over (P^2)^n} ,
\\
\widetilde{\cal J}_n &\equiv&
	\hbox{$\sum$}\!\!\!\!\!\! \int_{\{P\}} {p_0^2 \over (P^2)^{n+1}} ,
\\
\widetilde{\cal K}_n &\equiv&
	\hbox{$\sum$}\!\!\!\!\!\! \int_{\{P\}} {p_0^4 \over (P^2)^{n+2}} .
\end{eqnarray}
The specific sum-integrals that are needed are
\begin{eqnarray}
\widetilde{\cal I}'_0 &=& - {7 \pi^2 \over 360} T^4
	\left[ 1 + O(\epsilon) \right]\,,
\\
\widetilde{\cal I}_1 &=& - {1 \over 24} T^2
\left( {\Lambda \over 4 \pi T} \right)^{2 \epsilon}
\left[1 +  \left( 2 - 2 \log 2 + 2 {\zeta'(-1) \over \zeta(-1)} \right)
\epsilon
	+ O(\epsilon^2) \right]\,,
\\
\widetilde{\cal J}_1 &=& {1 \over 48} T^2
\left( {\Lambda \over 4 \pi T} \right)^{2 \epsilon}
\left[ 1 + \left( -2 \log 2 + 2 {\zeta'(-1) \over \zeta(-1)} \right) \epsilon
	+ O(\epsilon^2) \right] \,,
\\
\widetilde{\cal I}_2 &=& {1 \over (4\pi)^2}
\left( {\Lambda \over 4 \pi T} \right)^{2 \epsilon}
\left[ {1 \over \epsilon} + 4 \log 2 + 2 \gamma + O(\epsilon) \right]\,,
\\
\widetilde{\cal J}_2 &=& {1 \over 4 (4 \pi)^2}
\left( {\Lambda \over 4 \pi T} \right)^{2 \epsilon}
\left[ {1 \over \epsilon} + 2 + 4 \log 2 + 2 \gamma + O(\epsilon) \right]\,,
\\
\widetilde{\cal K}_2 &=& {1 \over 8 (4 \pi)^2}
\left( {\Lambda \over 4 \pi T} \right)^{2 \epsilon}
\left[ {1 \over \epsilon} + {8 \over 3} + 4 \log 2 + 2 \gamma
	+ O(\epsilon) \right]\,.
\end{eqnarray}

All of the two-loop sum-integrals that arise in the calculation
factor into the product of 2 one-loop sum-integrals.
Some of the three-loop sum-integrals factor into the product of
3 one-loop sum-integrals.  Others factor into the product
of a one-loop sum-integral and a two-loop sum-integral.  However,
 these sum-integrals all vanish, either because the one-loop sum-integral
is ${\cal I}_0 = 0$ or $\widetilde{\cal I}_0 = 0$, or because the
two-loop sum-integral vanishes:
\begin{eqnarray}
\hbox{$\sum$}\!\!\!\!\!\! \int_{PQ} {1 \over P^2 Q^2 (P+Q)^2} &=& 0 \,,
\label{a8}
\\
\hbox{$\sum$}\!\!\!\!\!\! \int_{\{P\}Q} {1 \over P^2 Q^2 (P+Q)^2} &=& 0 \,.
\end{eqnarray}
The remaining three-loop sum-integrals have the following forms:
\begin{eqnarray}
{\cal M}_{i,j} &\equiv&
\hbox{$\sum$}\!\!\!\!\!\! \int_{PQR}
  {1 \over P^2 Q^2 [R^2]^i [(P-Q)^2]^j (Q-R)^2 (R-P)^2} \;,
\\
\widetilde{\cal M}_{i,j} &\equiv&
\hbox{$\sum$}\!\!\!\!\!\! \int_{\{PQR\}}
  {1 \over P^2 Q^2 [R^2]^i [(P-Q)^2]^j (Q-R)^2 (R-P)^2} \;,
\\
{\cal N}_{i,j} &\equiv&
\hbox{$\sum$}\!\!\!\!\!\! \int_{\{PQ\}R}
  {1 \over P^2 Q^2 [R^2]^i [(P-Q)^2]^j (Q-R)^2 (R-P)^2} \;.
\end{eqnarray}
These sum-integrals can be evaluated analytically using methods
developed by Arnold and Zhai \cite{arnold-zhai}.
The specific integrals that are needed are
\begin{eqnarray}
{\cal M}_{0,0}
&=& {1 \over 24 (4 \pi)^2} T^4
\left( {\Lambda \over 4 \pi T} \right)^{6 \epsilon}
\left[ {1 \over \epsilon}
	+ {91 \over 15} + 8 {\zeta'(-1) \over \zeta(-1)}
	- 2 {\zeta'(-3) \over \zeta(-3)}
	+ O(\epsilon) \right] \;,
\\
\widetilde{\cal M}_{0,0}
&=& - {1 \over 192 (4 \pi)^2} T^4
\left( {\Lambda \over 4 \pi T} \right)^{6 \epsilon}
\left[ {1 \over \epsilon}
	+ {179 \over 30} - {34 \over 5} \log 2 + 8 {\zeta'(-1) \over \zeta(-1)}
\right.
\nonumber \\
&& \left.
\qquad \qquad \qquad \qquad \qquad \qquad \qquad \qquad
	- 2 {\zeta'(-3) \over \zeta(-3)} + O(\epsilon) \right] \,,
\\
{\cal N}_{0,0}
&=&  {1 \over 96 (4 \pi)^2} T^4
\left( {\Lambda \over 4 \pi T} \right)^{6 \epsilon}
\left[ {1 \over \epsilon}
	+ {173 \over 30} - {42 \over 5} \log 2 + 8 {\zeta'(-1) \over \zeta(-1)}
\right.
\nonumber \\
&& \left.
\qquad \qquad \qquad \qquad \qquad \qquad \qquad \qquad
	   - 2 {\zeta'(-3) \over \zeta(-3)}  + O(\epsilon) \right] \,,
\\
\widetilde{\cal M}_{1,-1}
&=& - {1 \over 192 (4 \pi)^2} T^4
\left( {\Lambda \over 4 \pi T} \right)^{6 \epsilon}
\left[ {1 \over \epsilon}
	+ {361 \over 60} + {76 \over 5} \log 2 + 6 \gamma
        - 4 {\zeta'(-1) \over \zeta(-1)}
\right.
\nonumber \\
&& \left.
\qquad \qquad \qquad \qquad \qquad \qquad \qquad \qquad
	+ 4 {\zeta'(-3) \over \zeta(-3)} + O(\epsilon) \right] \,,
\\
{\cal M}_{2,-2}
&=& {11 \over 216 (4 \pi)^2} T^4
\left( {\Lambda \over 4 \pi T} \right)^{6 \epsilon}
\left[ {1 \over \epsilon}
	+ {73 \over 22} + {12 \over 11} \gamma
        + {64 \over 11} {\zeta'(-1) \over \zeta(-1)}
\right.
\nonumber \\
&& \left.
\qquad \qquad \qquad \qquad \qquad \qquad \qquad \qquad
	- {10 \over 11} {\zeta'(-3) \over \zeta(-3)}
	+ O(\epsilon) \right] \,,
\\
\widetilde{\cal M}_{-2,2}
&=&  -{29 \over 1728 (4 \pi)^2} T^4
\left( {\Lambda \over 4 \pi T} \right)^{6 \epsilon}
\left[ {1 \over \epsilon}
	+ {89 \over 29} + {48 \over 29} \gamma - {90 \over 29} \log 2
        + {136 \over 29} {\zeta'(-1) \over \zeta(-1)}
\right.
\nonumber \\
&& \left.
\qquad \qquad \qquad \qquad \qquad \qquad \qquad \qquad
	- {10 \over 29} {\zeta'(-3) \over \zeta(-3)}
	+ O(\epsilon) \right] \,,
\\
{\cal N}_{2,-2}
&=& {1 \over 108 (4 \pi)^2} T^4
\left( {\Lambda \over 4 \pi T} \right)^{6 \epsilon}
\left[ {1 \over \epsilon}
	+ {35 \over 8} + {3 \over 2} \gamma - {63 \over 10} \log 2
        + 5 {\zeta'(-1) \over \zeta(-1)}
\right.
\nonumber \\
&& \left.
\qquad \qquad \qquad \qquad \qquad \qquad \qquad \qquad
	- {1 \over 2} {\zeta'(-3) \over \zeta(-3)}
	+ O(\epsilon) \right] \,.
\end{eqnarray}

\section{Integrals in the Effective Theory}
\setcounter{equation}{0}

The effective theory for the scale $g T$ is an Euclidean field theory
in 3 space dimensions. Loop diagrams involve integrals over 3-momenta.
It is convenient to introduce the notation $\int_p$ for these integrals.
We use dimensional regularization in $3-2\epsilon$ dimensions to
regularize both infrared and ultraviolet divergences.
We define the integration measure
\begin{equation}
  \int_p\;\equiv\;
  \left(\frac{e^\gamma \Lambda^2}{4\pi}\right)^\epsilon\,
  \int {d^{3-2\epsilon}p \over (2 \pi)^{3-2\epsilon}}\,.
\end{equation}
If renormalization is accomplished by the minimal subtraction of poles
in $\epsilon$, then $\mu$ is the renormalization scale in the
$\overline{\rm MS}$ scheme.  Below, we collect all the integrals that
are needed to calculate the contribution to the free energy
>from the momentum scale $gT$ to order $g^5$.

The nontrivial one-loop integrals that arise in the calculation have
the form
\begin{equation}
I_n \;\equiv\;
\int_p {1 \over [p^2 + m^2]^n} \;.
\end{equation}
The specific one-loop integrals that are needed are
\begin{eqnarray}
I'_0 & = & {m^3 \over 4 \pi}
\left( {\Lambda \over 2 m} \right)^{2 \epsilon}
\left[ {2 \over 3} + {16 \over 9} \epsilon
	+ O(\epsilon^2) \right] \,,
\label{Iprime}
\\
I_1 & = & {m \over 4 \pi}
\left( {\Lambda \over 2 m} \right)^{2 \epsilon}
\left[- 1 - 2 \epsilon + O(\epsilon^2) \right] \,,
\\
I_2 & = & {1 \over 4 \pi m}
\left( {\Lambda \over 2 m} \right)^{2 \epsilon}
\left[ {1 \over 2} + O(\epsilon^2) \right] \,.
\end{eqnarray}
In  (\ref{Iprime}), $I'_0$ denotes the derivative of
$I_n$ with respect to $n$ evaluated at $n=0$.

Some of the two-loop integrals reduce to products of one-loop integrals.
The remaining two-loop integrals have the following forms:
\begin{eqnarray}
J_n &\equiv&
\int_{pq} {1 \over (p^2 + m^2) \left[ q^2 + m^2 \right]^n (p-q)^2} \;,
\\
K_n &\equiv&
\int_{pq} {1 \over (p^2 + m^2) (q^2 + m^2) \left[ (p-q)^2 \right]^n} \;.
\end{eqnarray}
The specific two-loop integrals that are needed are
\begin{eqnarray}
J_1 & = & {1 \over (4 \pi)^2}
\left( {\Lambda \over 2 m} \right)^{4 \epsilon}
\left[ {1 \over 4 \epsilon} + {1 \over 2} + O(\epsilon) \right] \,,
\\
J_2 & = & {1 \over (4 \pi)^2 m^2}
\left( {\Lambda \over 2 m} \right)^{4 \epsilon}
\left[ {1 \over 4}  + O(\epsilon) \right] \,,
\\
K_2 & = & {1 \over (4 \pi)^2 m^2}
\left( {\Lambda \over 2 m} \right)^{4 \epsilon}
\left[ - {1 \over 8}  + O(\epsilon) \right] \,.
\end{eqnarray}

Some of the three-loop integrals reduce to the product of 3 one-loop
integrals or to the product a a one-loop integral and a two-loop integral.
The remaining three-loop integrals have the form
\begin{eqnarray}
M_{i,j} & \equiv &
\int_{pqr} {1 \over (p^2 + m^2) (q^2 + m^2) [r^2 + m^2]^i}
	{1 \over [(p-q)^2]^j (q-r)^2 (r-p)^2} \;,
\\
N_{i,j} & \equiv &
\int_{pqr} {1 \over (p^2 + m^2) (q^2 + m^2) ((q-r)^2 + m^2)
	((r-p)^2 + m^2)} {1 \over [r^2]^i [(p-q)^2]^j } ,
\\
L_{i,j} & \equiv &
\int_{pqr} {1 \over (p^2 + m^2) [(r-p)^2 + m^2]^i
	[q^2 + m^2]^j ((q-r)^2 + m^2)} {1 \over r^2 (p-q)^2 } .
\end{eqnarray}
These integrals are special cases of more general three-loop integrals defined
by Broadhurst \cite{broadhurst}:
\begin{eqnarray}
M_{i,j} &=&
m^{1-2i-2j} \left( {\Lambda \over m} \right)^{6 \epsilon}
\left( e^{\gamma \epsilon} \Gamma({3 \over 2} + \epsilon)
	\over (4 \pi) ^{3/2} \right)^3
B_M(1,j,1,1,1,i) \;,
\\
N_{i,j} &=&
m^{1-2i-2j} \left( {\Lambda \over m} \right)^{6 \epsilon}
\left( e^{\gamma \epsilon} \Gamma({3 \over 2} + \epsilon)
	\over (4 \pi) ^{3/2} \right)^3
B_N(i,j,1,1,1,1) \;,
\\
L_{i,j} &=&
m^{1-2i-2j} \left( {\Lambda \over m} \right)^{6 \epsilon}
\left( e^{\gamma \epsilon} \Gamma({3 \over 2} + \epsilon)
	\over (4 \pi) ^{3/2} \right)^3
B_N(1,1,1,1,i,j) \;.
\end{eqnarray}
Broadhurst derived recursion equations for the integrals $B_M$
and $B_N$ with general arguments which can be used to reduce
any of the integrals $M_{i,j}$, $N_{i,j}$ and $L_{i,j}$ to
the basic integrals $M_{0,0}$ and $N_{0,0}$, together with
simpler 1-loop and 2-loop integrals.
The specific integrals that are needed in our calculation are
\begin{eqnarray}
M_{0,0} & = & {m\over (4 \pi)^3}
\left( {\Lambda \over 2 m} \right)^{6 \epsilon}
\left[ - {1 \over 2 \epsilon} - 4 + O(\epsilon) \right] \,,
\\
M_{-1,1} & = & {m \over (4 \pi)^3}
\left( {\Lambda \over 2 m} \right)^{6 \epsilon}
\left[ {1 \over 4 \epsilon} + 2 + O(\epsilon) \right] \,,
\\
M_{-2,2} & = & {m \over (4 \pi)^3}
\left( {\Lambda \over 2 m} \right)^{6 \epsilon}
\left[ - {1 \over 4 \epsilon} - {3 \over 2} + O(\epsilon) \right] \,,
\\
M_{1,0} & = & {1 \over (4 \pi)^3 m}
\left( {\Lambda \over 2 m} \right)^{6 \epsilon}
\left[ {\pi^2 \over 12} + O(\epsilon) \right] \,,
\\
M_{0,1} & = & {1 \over (4 \pi)^3 m}
\left( {\Lambda \over 2 m} \right)^{6 \epsilon}
\left[ - {1 \over 8 \epsilon} + {1 \over 4} + O(\epsilon) \right] \,,
\\
M_{2,0} & = & {1 \over (4 \pi)^3 m^3}
\left( {\Lambda \over 2 m} \right)^{6 \epsilon}
\left[ - {1 \over 4} + {\pi^2 \over 24} + O(\epsilon) \right] \,,
\\
N_{0,0} & = & {m \over (4 \pi)^3}
\left( {\Lambda \over 2 m} \right)^{6 \epsilon}
\left[ - {1 \over \epsilon} - 8 + 4 \log 2 + O(\epsilon) \right] \,,
\\
N_{1,-1} & = & {m \over (4 \pi)^3}
\left( {\Lambda \over 2 m} \right)^{6 \epsilon}
\left[ 2 - 4 \log 2 + O(\epsilon) \right] \,,
\\
N_{2,-2} & = & {m \over (4 \pi)^3}
\left( {\Lambda \over 2 m} \right)^{6 \epsilon}
\left[ - 3 + 4 \log 2 + O(\epsilon) \right] \,,
\\
N_{1,0} & = & {1 \over (4 \pi)^3 m}
\left( {\Lambda \over 2 m} \right)^{6 \epsilon}
\left[ \log 2 + O(\epsilon) \right] \,,
\\
N_{2,-1} & = & {1 \over (4 \pi)^3 m}
\left( {\Lambda \over 2 m} \right)^{6 \epsilon}
\left[ {1 \over 3} - {1 \over 3} \log 2 + O(\epsilon) \right] \,,
\\
N_{2,0} & = & {1 \over (4 \pi)^3 m^3}
\left( {\Lambda \over 2 m} \right)^{6 \epsilon}
\left[ - {1 \over 24} - {1 \over 12} \log 2 + O(\epsilon) \right] \,,
\\
N_{1,1} & = & {1 \over (4 \pi)^3 m^3}
\left( {\Lambda \over 2 m} \right)^{6 \epsilon}
\left[ {1 \over 4} - {1 \over 4} \log 2 + O(\epsilon) \right] \,.
\end{eqnarray}
We also require the sum of the integrals $M_{1,-1}$
and $L_{1,-1}$, which is simpler to calculate
than the individual integrals:
\begin{eqnarray}
M_{1,-1} \;+\; L_{1,-1}
&=& - M_{0,0} \;+\; 2 I_1 J_1
\nonumber
\\
&=& {m \over (4 \pi)^3} \left( {\Lambda \over 2 m} \right)^{6 \epsilon}
\left[ 2 + O(\epsilon) \right] \,.
\end{eqnarray}
%


\newpage
\section*{Figure Captions}

\begin{enumerate}

\item
One-loop Feynman diagrams for the gluon self-energy.  Curly lines,
solid lines, and dashed lines represent the propagators of gluons,
quarks, and ghosts, respectively.

\item
Two-loop Feynman diagrams for the gluon self-energy.  The solid blob
represents the sum of the one-loop gluon self-energy diagrams shown
in Fig.~1.

\item
One-loop Feynman diagrams for the free energy.

\item
Two-loop Feynman diagrams for the free energy.

\item
Three-loop Feynman diagrams for the free energy.

\item
One-loop Feynman diagrams for the logarithm of the partition function
of EQCD.

\item
Two-loop Feynman diagrams for the logarithm of the partition function
of EQCD.

\item
Three-loop Feynman diagrams for the logarithm of the partition function
of EQCD.

\end{enumerate}


\newpage
\begin{thebibliography}{99}


\bibitem{g-p-y}
D.J.~Gross, R.D.~Pisarski, and L.G.~Yaffe,
  Rev.\ Mod.\ Phys.\ {\bf 53}, 43 (1981).

\bibitem{appelquist-pisarski}
T.~Appelquist and R.D.~Pisarski, Phys.\ Rev.\ {\bf D23}, 2305 (1981).

\bibitem{nadkarni-1}
S.~Nadkarni, Phys.\ Rev.\ {\bf D27}, 917 (1983).

\bibitem{landsman}
N.P.~Landsman, Nucl.\ Phys.\ {\bf B322}, 498 (1989).

\bibitem{eft}
E. Braaten and A. Nieto, Phys. Rev. {\bf D51}, 6990 (1995).

\bibitem{f-k-r-s}
K. Farakos, K. Kajantie, K. Rummukainen, and M. Shaposhnikov,
        Phys.\ Lett.\ {\bf B336}, 494 (1994);
	Nucl.\ Phys.\ {\bf B425}, 67 (1994); {\bf B442}, 317 (1995).

\bibitem{solution}
E.~Braaten, Phys. Rev. Lett. {\bf 74}, 2164 (1995).

\bibitem{linde}
A.D.~Linde, Rep.\ Prog.\ Phys.\ {\bf 42}, 389 (1979);
            Phys.\ Lett.\ {\bf B96}, 289 (1980).

\bibitem{polyakov}
E. Braaten and A. Nieto, Phys. Rev. Lett. {\bf 74}, 3530 (1995).

\bibitem{arnold-yaffe}
P. Arnold and L.G. Yaffe, Washington preprint UW/PT-95-06 (hep-ph 9508280).

\bibitem{mclerran}
B.A. Freedman and L.D. McLerran, Phys. Rev. {\bf D16}, 1147 (1977);
	{\bf D16}, 1169 (1977);
V. Baluni, Phys. Rev. {\bf D17}, 2092 (1978).

\bibitem{f-s-t}
J.~Frenkel, A.V.~Saa, and J.C.~Taylor, Phys.\ Rev.\ {\bf D46}, 3670 (1992).

\bibitem{arnold-zhai}
P. Arnold and C. Zhai, Phys. Rev. {\bf D50}, 7603 (1994);
	Phys. Rev. {\bf D51}, 1906 (1995).

\bibitem{coriano-parwani}
C.~Coriano and R.R.~Parwani, Phys.\ Rev.\ Lett.\ {\bf 73}, 2398 (1994);
R.R.~Parwani and C.~Coriano, Nucl.\ Phys.\ {\bf B434}, 56 (1995).

\bibitem{parwani-singh}
R.~Parwani and H. Singh, Phys. Rev. {\bf D51}, 4518 (1995).

\bibitem{parwani}
R.R.~Parwani, Phys.\ Lett.\ {\bf B334}, 420 (1994).

\bibitem{andersen}
J.O. Andersen, Oslo preprint  (hep-ph 9509409).

\bibitem{kastening-zhai}
B. Kastening and C. Zhai, Purdue preprint PURD-TH-95-03 (hep-ph 9507380).

\bibitem{fQCD}
E.~Braaten and A.~Nieto, Northwestern preprint NUHEP-TH-95-10 (hep-ph 9508406).

\bibitem{chapman}
S. Chapman, Phys. Rev. {\bf D50}, 5308 (1994).

\bibitem{nadkarni-2}
S. Nadkarni, Phys.\ Rev.\ Lett.\ {\bf 60}, 491 (1988);
	Phys. Rev. {\bf D38}, 3287 (1988).

\bibitem{shuryak}
E. Shuryak, J.E.T.P. {\bf 47}, 212 (1978).

\bibitem{kapusta}
J.I. Kapusta, Nucl. Phys. {\bf B148}, 461 (1979).

\bibitem{broadhurst}
D.J. Broadhurst, Z. Phys {\bf C54}, 599 (1992).

\bibitem{rebhan}
A. Rebhan, Phys. Rev. {\bf D48}, R3967 (1993);
	Nucl. Phys. {\bf B340}, 319 (1994);
E. Braaten and A. Nieto, Phys. Rev. Lett. {\bf 73}, 2402 (1994).

\bibitem{BLM}
S. Brodsky, G.P. Lepage and P. Mackenzie, Phys. Rev. {\bf D28}, 228 (1983).

\end{thebibliography}
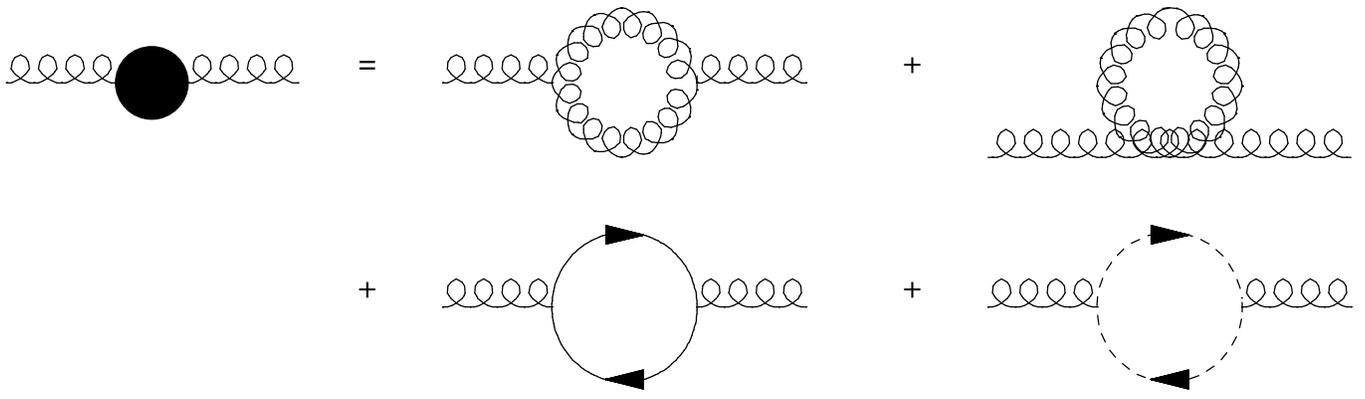
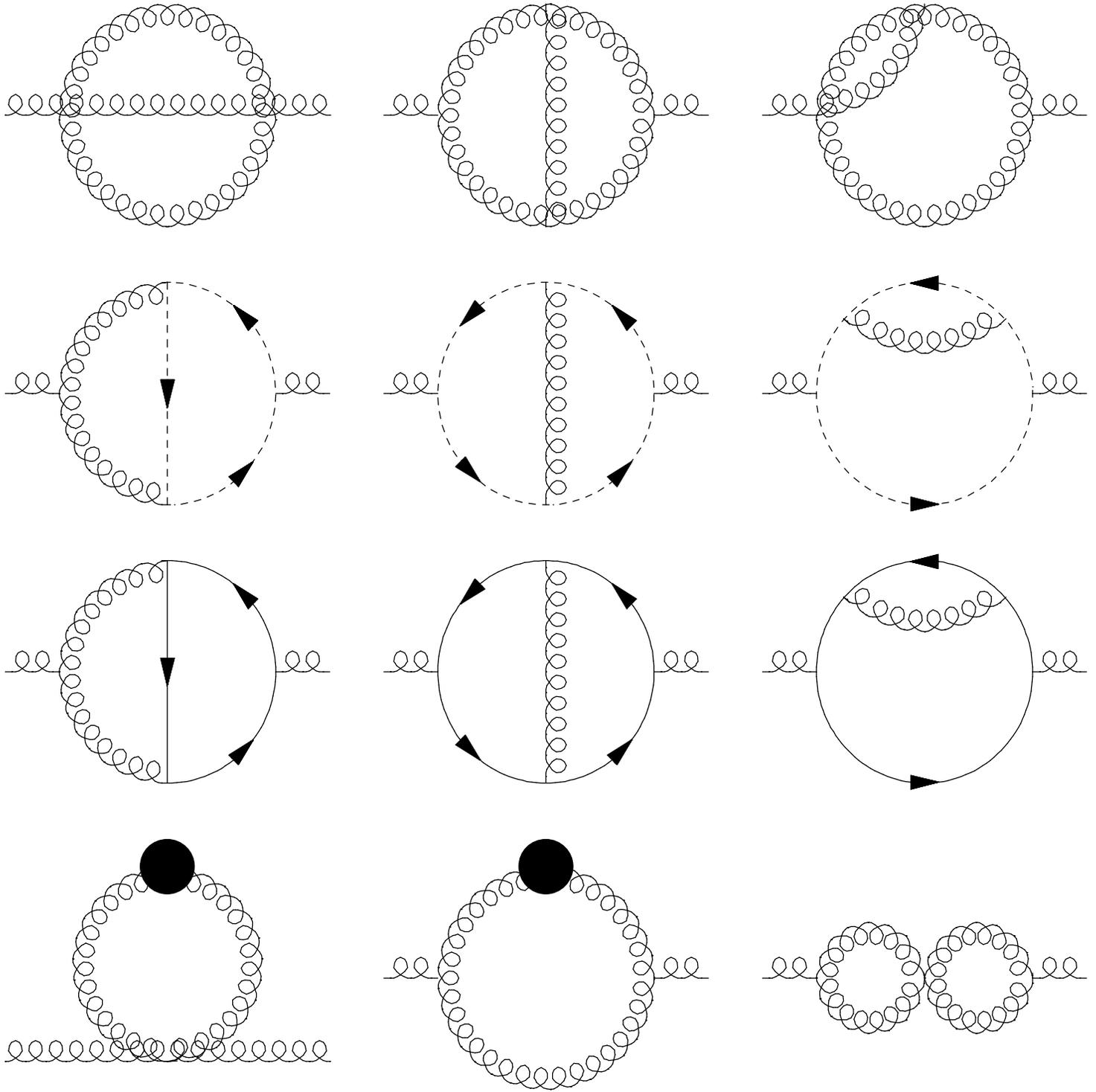
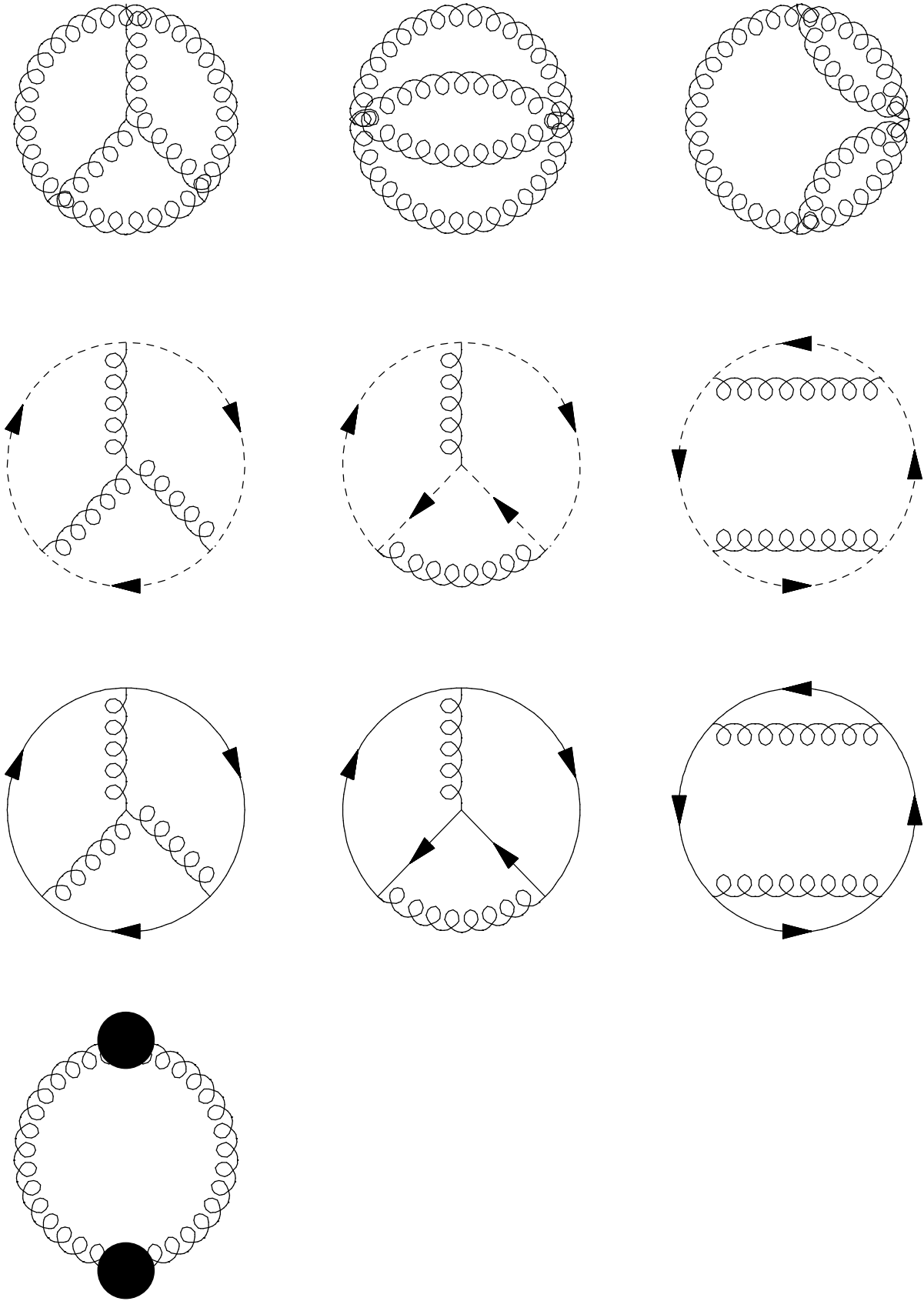
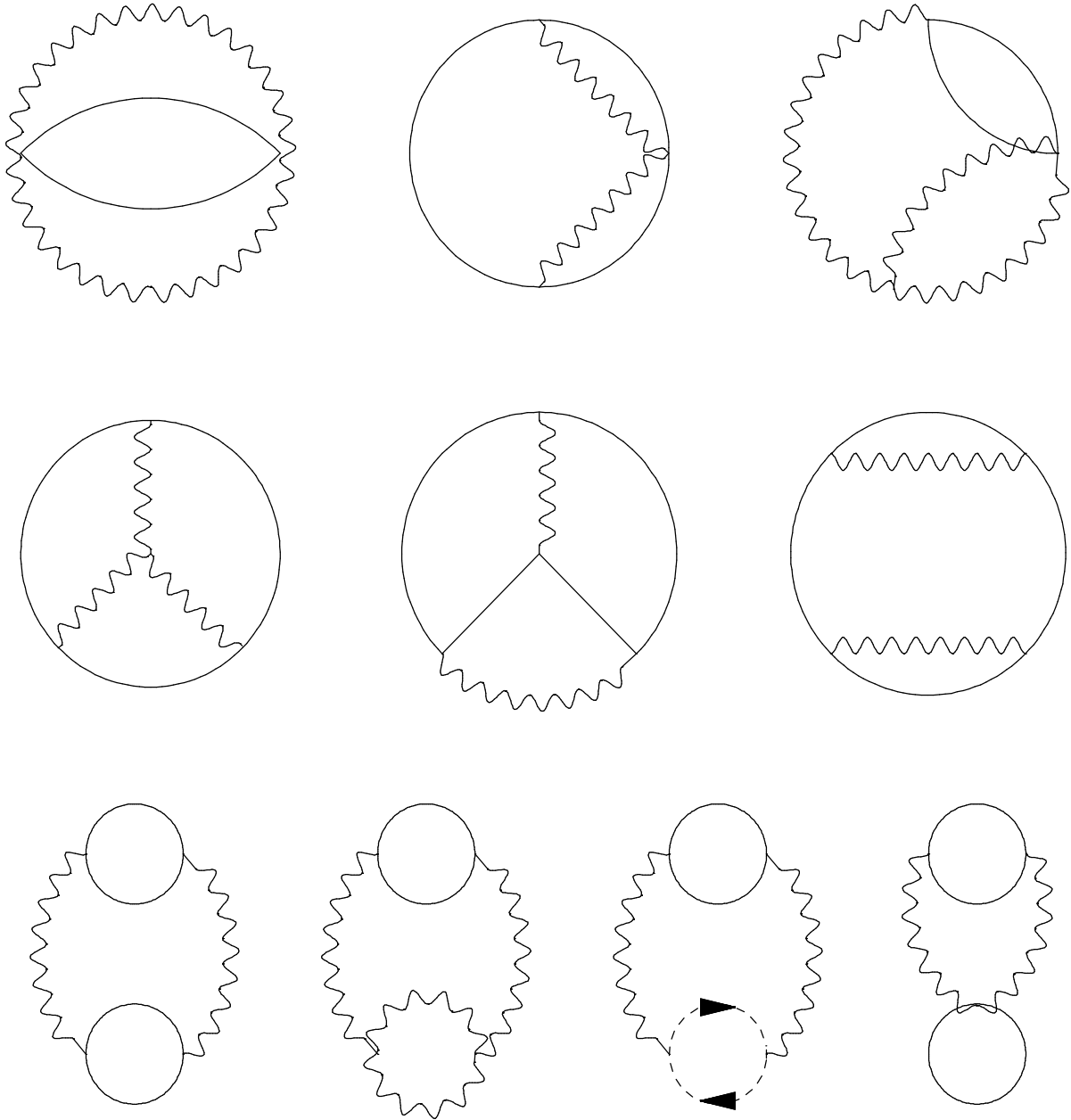
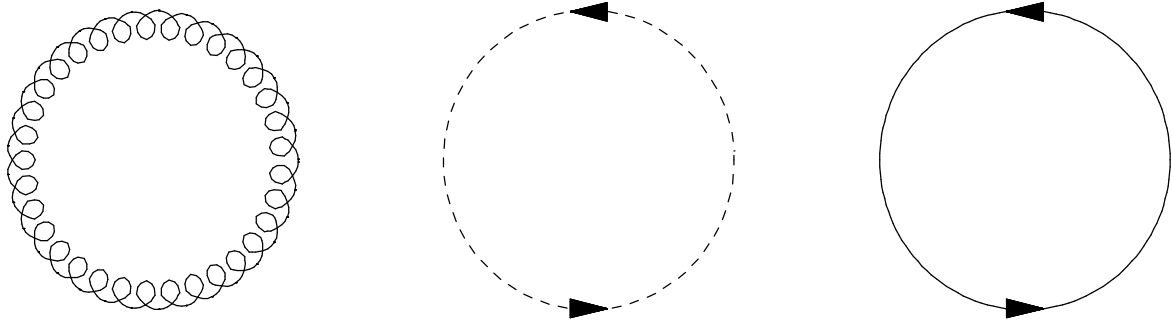
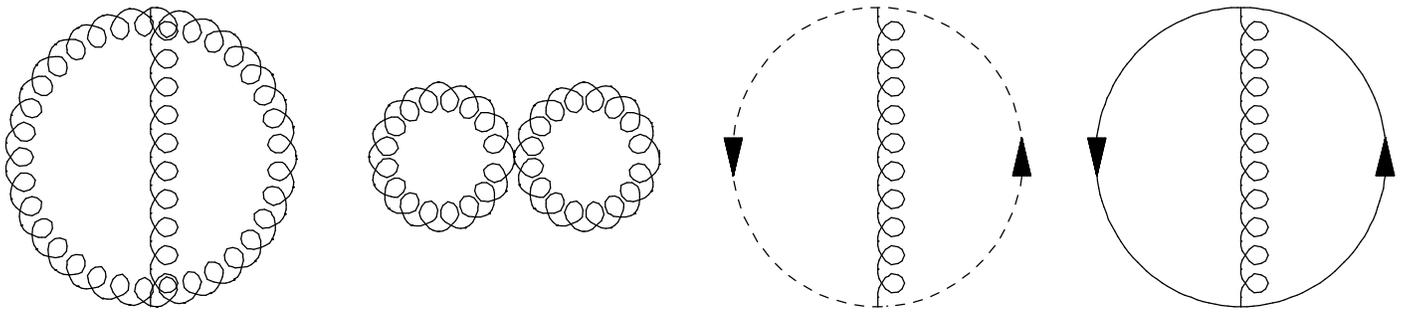
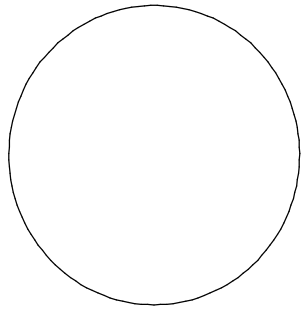
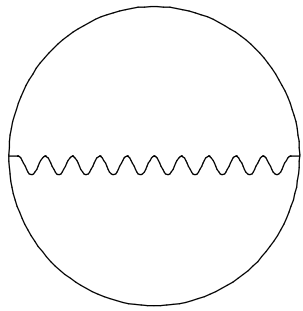
\end{document}